\documentclass{article}
\usepackage[a4paper, total={6in, 10in}]{geometry}
\usepackage[utf8]{inputenc}
\usepackage{cite}
\usepackage{hyperref}
\usepackage{amsmath}
\usepackage{xr}
\usepackage{graphicx}
\usepackage{xcolor}
\usepackage{float}
\DeclareUnicodeCharacter{2061}{}

\title{Role of mush complex viscosity in modulating axial topography in mid-oceanic ridges}
\author{Joyjeet Sen\footnote{senjoyjeet@gmail.com}, Shamik Sarkar\footnote{shamiksrakar@gmail.com}, Nibir Mandal\footnote{nibir.mandal@jadavpuruniversity.in}\\
Department of Geological Sciences, Jadavpur University,\\
Kolkata 700032, India\\
}
\date{}

\begin{document}
\maketitle

\begin{abstract}
This article exploits the interaction dynamics of the elastic oceanic crust with the underlying mush complexes (MC) to constrain the axial topography of mid-ocean ridges (MORs). The effective viscosity ($\mu_{eff}$) of MC beneath MORs is recognized as the crucial factor in modulating their axial high versus flat topography. Based on a two-step viscosity calculation (suspension and solid-melt mixture rheology), we provide a theoretical estimate of $\mu_{eff}$ as a function of melt suspension characteristics (crystal content, polymodality, polydispersity and strain-rate), and its volume fraction in the MC region. We then develop a numerical model to show the control of $\mu_{eff}$ on the axial topography. Using an enthalpy-porosity-based fluid-formulation of uppermost mantle the model implements a one-way fluid-structure interaction (FSI) that transmits viscous forces of the MC region to the overlying upper crust. The limiting non-rifted topographic elevations (-0.06 km to 1.27 km) of model MORs are found to occur in the viscosity range: $\mu_{eff}$ = $10^{12}$ to $10^{14}$ Pa s. The higher-end ($10^{13}$ to $10^{14}$) Pa s of this spectrum produce axial highs, which are replaced by flat or slightly negative topography as $\mu_{eff} \leq 5\times 10^{12}$ Pa s. We discuss a number of major natural MORs to validate the model findings. 
\end{abstract}

\section{Introduction}


Many mid-ocean ridges (MORs) evolve with complex 3D axial topography, which is hard to explain with standard tectonic models. Their spatially varying axial topography, such as high, flat or valley, is generally attributed to the spreading rate \cite{Small1998,Sim2020}, the magma availability \cite{Sinton1992,Keller2017,Mandal2018}, in a particular ridge-segment, and upper crustal faulting \cite{Buck2005}. However, these contrasting axial morphologies are often found in MORs, e.g., South-East Indian Ridge (SEIR), where the spreading rate shows practically no variations \cite{Carbotte2016}, and ultra-slow South-West Indian Ridges (SWIR) displaying typical axial valley topography, where they have large magma availability \cite{Jian2017}. A direction of MOR studies explains the rift morphology as a product of the two competing processes- tectonic and magmatic, conceived as horizontal spreading and dike opening, respectively \cite{Buck2005,Liu2018}. A non-dimensional parameter, called the M factor (a ratio of the dike intrusion to plate-spreading driven widening rates), has been used to reproduce the axial structures in numerical models. M $=$ 1, i.e., a condition of dike intrusion rate to completely balance with the plate spreading rate, gives rise to an axial high, whose height depends on the magma density. In contrast, M $<$ 1, i.e., a condition of less effective diking than the spreading rate, yields faulted axial valley \cite{Buck2005,Ito2008}. Some studies have shown the axial morphology as a function of the extension rate and inherent short-wavelength seafloor heterogeneities (e.g., \cite{Small1998}). However, their interpretation faces disagreement with the Mid-Atlantic ridge model, which proposes the magma supply as a critical factor in determining the axial morphology \cite{Liu2018}. Although these models well integrate the axial morphological spectrum by a single factor- M, the modelling approach does not account for sub-crustal melt processes. It is noteworthy that many recent MOR studies demonstrated how the latter could significantly control the MOR evolution \cite{Carbotte2016,Martinez2020}, albeit a comprehensive model is still unavailable. Our present article aims to bridge this gap, treating the axial morphology in the thermo-mechanical framework of an ideal three-dimensional melt upwelling system, where divergence force components act along and across the ridge axis. This modelling approach allows us to investigate the extent of magmatic control on 3D axial morphology.\par

While emphasizing magmatic roots, several workers considered magma buoyancy as the principal factor to elucidate the origin of axial-high topography \cite{Buck2001,Wilson1992}.\cite{Eberle1998} provided a condition of the sub-ridge viscosity distribution required for buoyancy-driven axial high topography. On the other hand, \cite{Morgan1987} predicted that mantle viscosities beneath the ridge must be at least two orders higher than the generally accepted values to form an axial valley. \cite{Sleep1979} also indicated viscosity as the key factor, but it is ultimately the plate velocity to regulate the sub-crustal density or viscosity that determine the axial morphology. Here, the most critical question is – how the plate velocity regulates the sub-crustal viscosity? \cite{Choi2010} showed from a 2D numerical model that the mantle viscosity at shallow depths ($<$ 20 km) beneath the ridge should be low ($\sim 10^{18}$ Pa s) to form axial high topography, but it should be high enough ($\sim 10^{21}$ Pa s) to form a low axial relief. According to their model, high-viscosity melt flows lower the hydrostatic pressure beneath the ridge, reducing the melt upwelling height. However, none of these studies explicitly accounts for the viscosity effect of sub-crustal melt-rich zones on the axial morphology.\par

\begin{figure}[h]
\includegraphics[width=\textwidth]{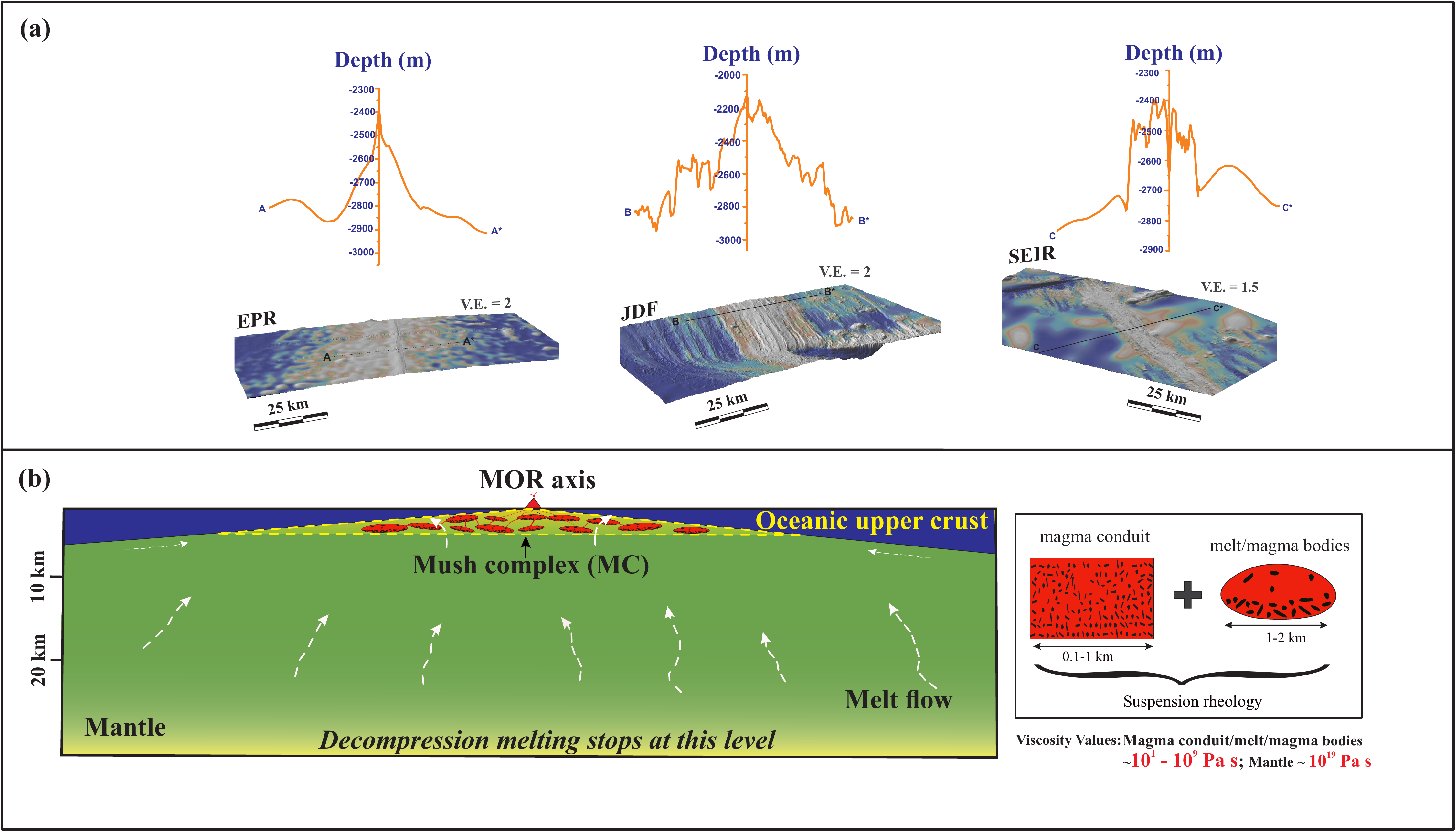}
\caption{(a) Bathymetric profiles across the East Pacific Rise (EPR), Juan De Fuca (JDF) and South-Eastern Indian Ridge (SEIR). They show high (EPR-AA*), moderately high (JDF-BB*) and plateau dominated (SEIR-CC*) ridge-axis topography, respectively. Data source: GeoMapApp (http://www.geomapapp.org/)/CC BY. (b) A conceptual cartoon diagram of the sub-ridge melt/magma settings considered for the topographic modelling in this study. The magma scale (Table\ref{ Table 1}) covers narrow melt conduits and magma bodies containing suspended crystals. The mush complex (MC) represents a distinct zone consisting of melt bodies and conduits within a high-viscosity host rock matrix.}
\label{Fig1}
\end{figure}

The problem of sub-crustal melt transport mechanisms has recently rejuvenated the MOR research in new directions \cite{Sparks2019, Edmonds2019, Carbotte2020, Carbotte2021}. It is now evident that melts start to localize in discrete zones during their ascent that eventually mediates for a heterogeneous magma supply to the ridge axes. Earlier numerical models \cite{Sarkar2014,Mandal2018} showed melt fraction as a function of spreading rates, suggesting that the melt fraction is substantially reduced from fast- to slow-spreading ridges. Secondly, the melt upwelling processes participate in solidification at the shallow level to form isolated mushy bodies, as reported by many earlier workers \cite{Sinton1992,Smith1992,Singh2006}. The crystal content in the mushy melts can largely vary depending on the degree of crystallization, and their varying relative volume ratios would determine the viscosity of the melt-bearing sub-ridge regions. \cite{Braun2000} provided a depth-wise viscosity profile based on melt content ($\sim$ 3\%), dehydration, and grain boundary sliding. This model predicts an increase in overall viscosity with height, mainly due to water extraction during partial melting. However, later experimental studies suggested that such dehydration can hardly affect viscosity at shallower depths as partial melting generally ceases to occur at a deeper level \cite{Hirth2003}. The olivine rich high-fluid channels in subcrustal magma mush at a shallower depth \cite{Kelemen2000} indicates crystal transport as suspension. A detailed viscosity analysis of the sub-crustal regions containing crystal-bearing melts beneath MORs, especially in view of the axial morphology, is yet to be fully explored.\par
This article introduces a novel approach to model sub-crustal/lower-crustal (hereafter sub-crustal) viscosity and offers a viscosity-based explanation for the axial morphologies: highs and flat topography of MORs, e.g., East Pacific Rise, Juan du Fuca, and South East Indian Ridge (Figure \ref{Fig1}a).In the first step, we provide a series of systematic calculations of the effective viscosity of mush complex (MC) beneath the ridge axes. The mush complex (MC) is defined here as a constitution of crystal-bearing-melts (with largely varying crystal contents) and host rocks \cite{Sparks2019}. Our calculations consider the following parameters: the process-times, spatial magnitudes, and the constitution of sub-crustal materials (see the concept diagram, Figure \ref{Fig1}b and Section 2). We then develop a three-dimensional fluid-structure interaction (FSI) approach to model the mechanical connection between the MC and the overlying crust at a mid-oceanic ridge. The FSI model allows us to investigate how the viscosity of the sub-crustal mushy region can modulate the flat versus high MOR axial topography.

\section{Sub-crustal mush complexes: viscosity modelling}

\subsection{Mush complex in MOR settings}

\begin{figure}[h]
\includegraphics[width=\textwidth]{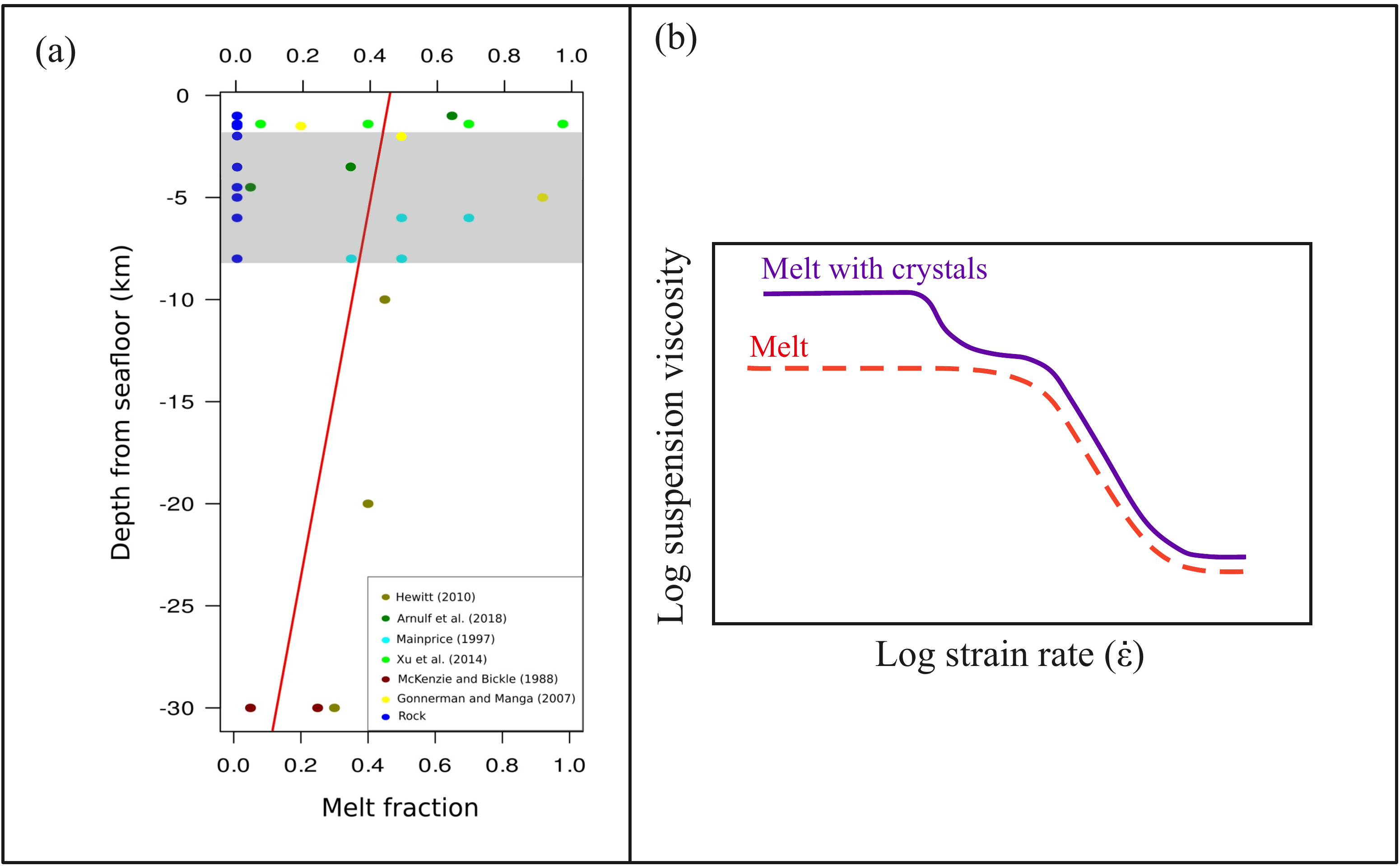}
\caption{(a) Variation of melt fractions with depth (plots based on available data in literature). Except Hewitt \cite{Hewitt2010} and McKenzie and Bickle \cite{McKENZIE1988}, all the data are taken from axial melt lenses, magma chambers and melt conduits. In the plot, these data points are complemented with rocks (1\% melt, \cite{Sparks2019}), marked in deep blue. Red straight line shows the overall regression trend. Shaded area delineates the depth range (2-8 km) of evolved MMC, where the average melt suspension fraction in MC is 0.3 - 0.4. (b) The plot shows the variation of suspension viscosity as a function of strain rate and characteristics of the melt suspension (modified from \cite{Gonnermann2007}).}
\label{Figure S1}
\end{figure}

At mid-oceanic ridges, the ascending melts produced by decompression melting $-$ at a depth of around 40 km $-$ focus to the ridge axis, forming large ($\sim$ 30 km wide) melt-rich regions \cite{McKENZIE1988}. Seismic imaging and theoretical estimates indicate a wide variation in their partial melt content ($\sim$ 10 - 70\% at shallower levels, $\leq$ 10 km and 5 – 25\% at deeper levels, $\sim$ 30  km) from one ridge to the other or different segments within the same ridge \cite{McKENZIE1988,Hewitt2010,Bergantz2015,Sauter2016}. It is vital to assess how such variations in melt content can influence the mechanical strength of sub-crustal melt-rich regions (MC) at shallow depths and modulate the first-order ridge-axis topography. In submarine systems, the temperature calculated for the critical depth of partial melting cessation constrains the amount of available melt in the subcrustal MC system \cite{McKENZIE1988}. However, the melts ascend upward with a complex 3D pattern of their paths, determined by coupled convection-solidification processes \cite{Sarkar2014,Mandal2018,Zhang2015}. The volume fraction of melt-crystal aggregates goes up \cite{Gonnermann2007,Hewitt2010}as subcrustal magma bodies form at mid-oceanic ridges. The plot shows a linear regression of the average melt fractions with depth (Figure \ref{Figure S1}a). There can be large variations from the linear average at subcrustal regions due to significant spatial variations in the magma pool populations and their fractional crystallization beneath MORs(Figure \ref{Figure S1}b).\par

Geophysical signatures, such as lower seismic velocities and high attenuation suggest the occurrence of mushy zones beneath mid-ocean ridges \cite{Arnoux2019,Wilson1992}, containing suspension-rich melt bodies (super solidus) as well as subsolidus host rocks \cite{Sparks2019}. Based on such sub-ridge mush-melt patterns reported in the literature \cite{Lin1989, Hebert2010, Keller2017, Sim2020}\cite{McKENZIE1988}\cite{McKENZIE1984}, we consider a mechanically distinct zone, mush complex (MC), treated as a continuum to implement the dynamic and kinematic coupling between the underlying mantle and the overlying elastic crust \cite{Carbotte2021, Zhang2015}. It is now a well-established fact that ascending mushy melts encounter the lithospheric base that acts as a melt barrier and forces the melts to focus into the ridge axis, forming a distinct melt-rich regime within the host rocks \cite{Lin1989, Hebert2010, Keller2017, Sim2020}. From gravity anomaly data, Lin and Parmentier\cite{Lin1989} detected a low-density zone at the base of the lithosphere at EPR. Mckenzie and Bickle\cite{McKENZIE1988} discussed the occurrence of underlying hot sheets at the decoupling zone between the circulating mantle and the spreading plates. On the other hand, many geophysical studies found sub-crustal melt lenses, 1-2 km wide and 100 m thick, containing 30-40\% crystals as suspension, in several ridges. The lenses are extended horizontally up to 15-20 km with their melt content decreasing to 30\%, forming spatially extensive mushy regions \cite{Singh2006, Canales2005, Toomey2007}. Singh \emph{et al.}\cite{Singh2006} recognized 3-4 km thick axial magma chambers at a depth of 3 km in slow-spreading, magma-rich Lucky strike. Fast and intermediate spreading ridges are reported to have magma bodies at shallower depths, $\sim$ 2.4 km in JdF \cite{Canales2005} and $\sim$ 1.8 km in EPR \cite{Detrick1987}, where their maximum thickness is $\sim$ 4 to 6 km \cite{Detrick1987, Kent1993}. In contrast, slow ridges generally lack such distinct magma bodies, but have mushy regions (low-velocity zones) at the crustal base. Dunn \emph{et el.}\cite{Dunn2005} reported a 6 km thick mushy zone of sparse melt channels at a depth of 4 km at MAR ($35^{\circ}$N). Besides melt pockets, which are prevalent in fast-spreading ridges, discrete melt channels in the lower crust and sub-crustal axial zones \cite{Dunn2005, Canales2005} also constitute a typical feature (low-velocity mushy regions) at the crustal base, which is also considered as a part of the MC \cite{Sinton1992}.\par

Based on the available reports on axial melt lenses and melt-rich bodies beneath mid-ocean ridges, the vertical extent of the mush complex (MC) was chosen in the present model in sub-crustal regions and in the uppermost mantle. The MC was allowed to evolve with progressively deforming overlying elastic upper crust under basal stresses that eventually decreased the axial depth and increased the MC thickness.  The MC is modelled with a triangular cross-section, describing an along-axis prismatic area in the lower crust, with a maximum thickness of 4 km beneath the ridge axis (Figure \ref{Fig1}b). The upper crust (i.e., solid elastic crust) at the axis is chosen 4 km thick in the initial model setting, where the MC vertically covers the lower crust and a part of the topmost mantle region (detailed in, Figure \ref{Fig3}). The reason for choosing a larger MC depth, as compared to the available data in our initial model is that the solid crust progressively thins by elastic strains during the simulation. For example, the initial crustal thickness at the axis is reduced by 2 km to finally set the MC depth and thickness at 2 km and 6 km, respectively \cite{Sinton1992}.

\subsection{ Melt-viscosity modelling: parametric considerations }

The viscosity of melts and magmas at shallow depths is historically modelled within a framework of suspension rheology, following the landmark work of \cite{Einstein1906}. However, natural suspensions show viscous behaviour more complex than that predicted from Einstein’s theory. The complexity originates primarily from the effects of additional factors, such as packing and shapes of solid particles in suspensions. The packing of solid components in the liquid phase is an influential factor to modify the effective viscosity of a mixture under the same solid volume fraction \cite{Krieger1959}. The packing vis-a-vis viscosity, depends significantly also on the solid particle size distribution in the liquid. For example, a bimodal size distribution with increasing size ratios up to a threshold point lowers the effective viscosity \cite{Chong1971}. Chang and Powel \cite{Chang1994} showed that the viscosity of a suspension decreases initially with increasing smaller particle volume fraction and then increases monotonously after reaching a critical volume fraction. Liquid suspensions attain their maximum packing fraction in the case of multimodal size distributions. Such multimodal (trimodal and tetramodal) particle packing increases viscosity higher than those for bimodal and unimodal distributions \cite{Klein2018}. This study also suggests that packing with different particle sizes yields a polydispersion effect on the bulk viscosity of suspensions for the same particle volume fraction. Polydispersity allows smaller particles to pack more closely by forming layers between larger particles or by occupying the void spaces between larger particles \cite{Desmond2014}. Such polymodal and polydisperse particle (heterogeneous packing) distributions can thus significantly enhance the bulk viscosity of melt suspensions in a mushy region.\par

\begin{figure}[h]
\includegraphics[width=\textwidth]{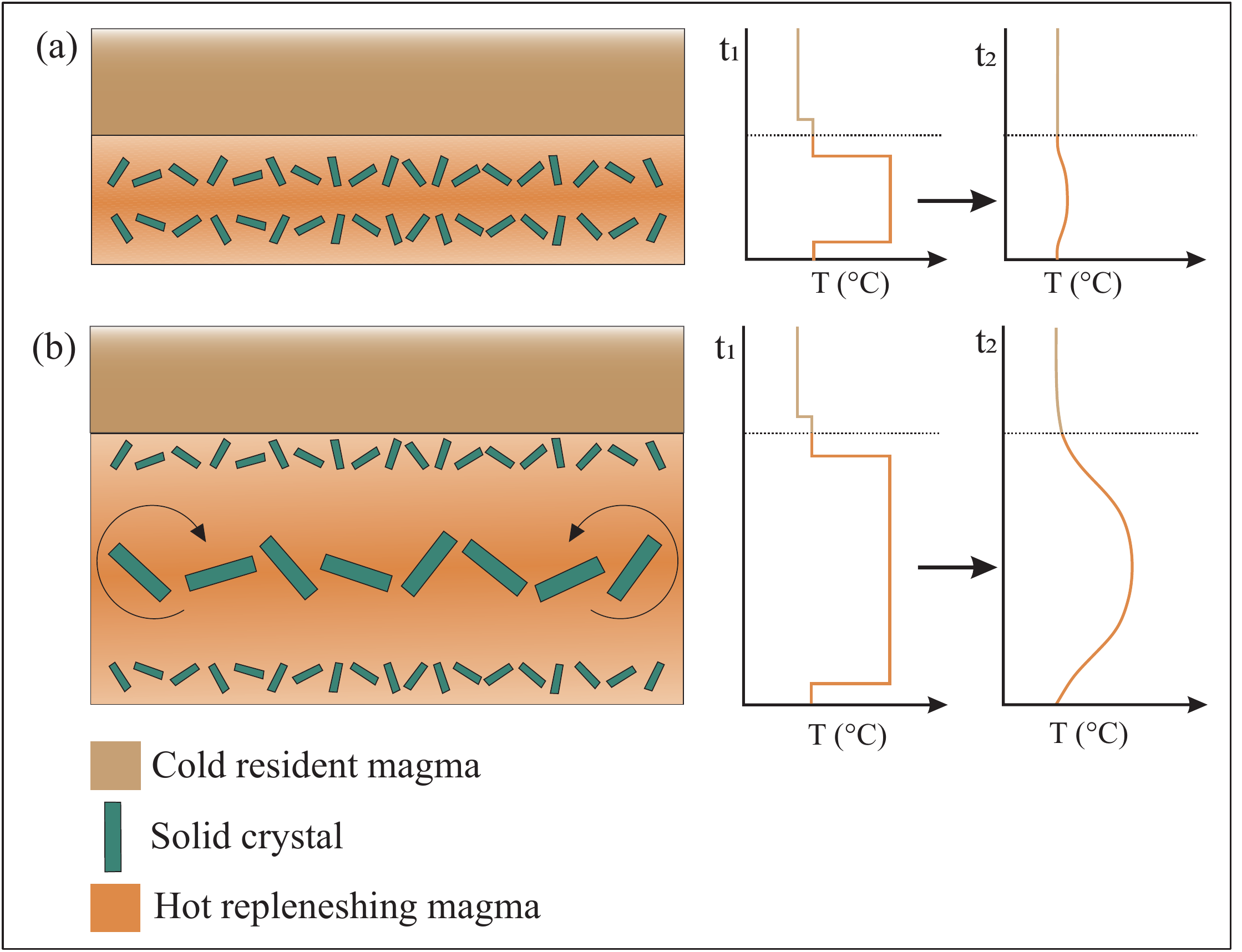}
\caption{A cartoon presentation of two different replenishment dynamics in the subcrustal magma-chambers (adapted from \cite{MARTIN2006}). (a) Volumetrically small and slow magma replenishment. In this case a thin layer of groundmass crystallization forms, and the thermal gradient between the hot emplaced magma and the host mafic magma is dissipated and reduced. (b) Fast and large magma replenishment. In this case, the hot magma disperses into a magma chamber like fountains before crystallizing to form enclaves. Moreover, a large volume of replenishing magma causes a thicker hot layer, which sustains the thermal gradient for a longer period with convective churning and multimodal crystal enclaves.}
\label{Figure S2}
\end{figure}

Petrological and compositional data indicate mineral phases, e.g., olivine, plagioclase and clinopyroxene can crystallize in partially molten zones at successive stages during the melt ascent. However, their depth correlation becomes weak with decreasing spreading rate \cite{Klein1987,langmuir1992}. Volcanic studies suggest that the polydispersity characteristics of mushy melts and their variations hold a connection with the spreading rates at MORs. Mushy melts can strongly differ in their crystal and bubble contents depending upon the spreading rates \cite{Edmonds2019,Kavanagh2018}. Slow-spreading ridges generally show larger compositional variations in erupting lavas than the fast-spreading ridges. Higher degrees of compositional homogenization in the fast-spreading ridges are commonly attributed to the magma chamber processes \cite{ohara1981}, in contrast to slow-spreading ridges, e.g., the Mid Atlantic Ridge (MAR), which are devoid of large, stable magma chambers and involves fractional melting throughout the whole melting regime to produce melts with a strong compositional variability \cite{Shimizu1998}. It is noteworthy that magma chamber processes act as potential sites for hot mafic magma replenishment into the cold silicic resident magmas (Figures \ref{Figure S2}a and \ref{Figure S2}b). Several workers have reported mafic enclaves from volcanic rocks as an evidence of the magma replenishing process (Figure \ref{Figure S2}a) \cite{MURPHY2000,Coombs2004,MARTIN2006}.

The effects of replenishment dynamics on magma chamber processes depend mainly on the magma flow conditions. Turbulent fountain-type injections produce dispersed spherical enclaves of crystals, whereas a non-turbulent slow influx of hot magma (Re $\le$ 400) (Figure \ref{Figure S2}) results in ponding of groundmass crystals at the magma-chamber base \cite{Coombs2004,MARTIN2006}. Martin \emph{et al.} \cite{MARTIN2006} also suggested that the more replenishing magma volume flux, the larger olivine phenocrysts would form, resulting in convection-driven churning within the magma chamber due to a longstanding steeper temperature gradient (Figure \ref{Figure S2}b). However, recent geochemical studies focus on the general mushy environment as a host of crystallization and magma mixing \cite{Bachmann2016}. Geochemical investigations of crystal-bearing enclaves within erupted lavas in volcanic settings allow us to recognize several factors controlling shallow magmatic environments beneath mid-oceanic ridges:  a) occurrence of magma chamber, b) melt-upwelling velocity, and c) volume in the mushy regions. These factors determine whether magma-dominated, volatile environments can significantly influence the dispersion and multimodality characteristics of solid suspensions in hot magmas. In contrast, magma-poor mushy regions, consisting of small and unstable non-convecting magma chambers, can form dominantly monodispersed microlith ponds.\par
 
Petrogenetic analysis of a recent East Pacific Rise (EPR) eruption suggests the presence of sub-ridge crystal-bearing mushy regions, which are thought to be a product of consistent replenishment (Figures \ref{Figure S2}a and \ref{Figure S2}b) by evolved magmas from a deeper level source \cite{Goss2010}. Rubin \emph{et al.},\cite{Rubin2001} reported homogeneity in lava basalts from fast ridges [Heterogeneity Index (HI) $\sim$ 1.6 for spreading rate $>10$ cm/yr], but strong heterogeneity from slower ones [HI $\sim$ 3 for spreading rate $<$ 4 cm/yr]. They accounted for the relative thermal stability to explain the higher degree of homogeneity in the fast ridges. Phyric basalts containing mushy zone crystals suggest the injection of primitive magmas. However, the petrological estimates of mid-ocean ridge basalt (MORB) at MAR point to larger volumes of aphyric basalt, representing a collection of aphyric magmas within a mushy zone at shallow depths. The magmas originated from convection-assisted melt segregation. Lange \emph{et al.} \cite{Lange2013} calculated mush viscosity \cite{giordano2008} as a function of plagioclase phenocryst content. According to their estimate, the viscosity of melt suspensions can increase by eight times with an increase in small-size ($\sim$ max. 10 mm) plagioclase phenocryst fraction, where the crystallinity is 20\%. Their analysis predicts the maximum size of olivine phenocrysts in erupting plagioclase-ultraphyric-basalts (PUB) in the range of 1 to 3 mm. They also explain the presence of PUB selectively in slow and intermediate ridges as a consequence of olivine phenocryst segregation in conduits during the magma ascent rather than in the magma chambers. Lange \emph{et al.} \cite{Lange2013} hypothesized that during the ascent of melt-crystal aggregates through conduits, the melt to crystal ratio is high, and their bulk viscosity is thereby low. However, it is hard for low-viscosity magmas to transport crystals without segregation in the conduit. Going by these arguments for the petrogenesis of plagioclase-phyric basalts, we infer that the bulk viscosity of magmas at the time of their ascent through conduits must be low in cases of slow and moderately spreading ridges, allowing extensive crystal segregation to produce monomodal crystal packing with little or no polydispersity. In contrast, the melt suspension viscosity in fast-spreading ridges, or ridges with prominent magma chambers, should be high due to greater polydispersity and polymodality even the melt volume fraction is relatively large. The polydispersity variation in sub-ridge magmatic processes is thus an influential factor to modulate the melt suspension viscosity. Based on the preceding discussions of melt characteristics, we thus consider the packing factors in the viscosity analysis of crystal-bearing melts in mushy regions.

\begin{center}
\begin{table}[h]
\caption{Model Parameters used in determining the viscosity scale}
\label{ Table 1}
\begin{tabular}{llll}
\hline 
Domain \hspace{30mm} & Properties \\ 
\hline \\ 
{\bf Lava Scale } & Conduit length ($l_c$) = 0.1 km; Conduit diameter ($d_c$) = 0.1 km;\\
&Transmitted volume ($v_c$) = 0.01 $\text{km}^3$; Strain rate ($\dot{\varepsilon}$)= $10^{-5} s^{-1}$\\ 
&Transmitting time ($t_c$) = 5 hrs\\ \\ 
{\bf Magma Scale } & Conduit length ($l_c$) = 1 km; Conduit diameter ($d_c$) = 0.1 km;\\
&Transmitted volume ($v_c$) = 0.01 $\text{km}^3$; Strain rate ($\dot{\varepsilon}$)= $10^{-9} s^{-1}$ \\ 
&Transmitting time ($t_c$) = 13 yrs\\ \\ 
{\bf Lava Scale } & Conduit length ($l_c$) = 10 km; Conduit diameter ($d_c$) = 1 km;\\
&Transmitted volume ($v_c$) = 0.01 $\text{km}^3$; Strain rate ($\dot{\varepsilon}$)= $10^{-14} s^{-1}$ \\ 
&Transmitting time ($t_c$) = 100 hrs\\ \\
\hline
\end{tabular}
\end{table}
\end{center}

\subsection{Mush complex (MC) viscosity: a two-step calculation }

We calculate the viscosity of mush complex (MC) in two steps: 1) viscosity calculation of crystal-bearing melts, based on the theory of suspension rheology, called \emph{melt suspension}, and 2) viscosity of MC (host rock + melt suspensions), based on the theory of \emph{two-phase fluid mixtures}. To calculate the viscosity of a crystal-melt aggregate, we assume the solid (crystals) component as a suspension in the liquid matrix, as described in the preceding section. Again, the solid part in suspension is treated as rigid particles and the liquid part as a continuous, viscous medium. The calculation is carried out using the equations and calculations found in the literature for erupted lavas. Figure \ref{Fig1}b provides a cartoon diagram to show its conceptual framework. 

\begin{figure}[H]
\includegraphics[width=\textwidth]{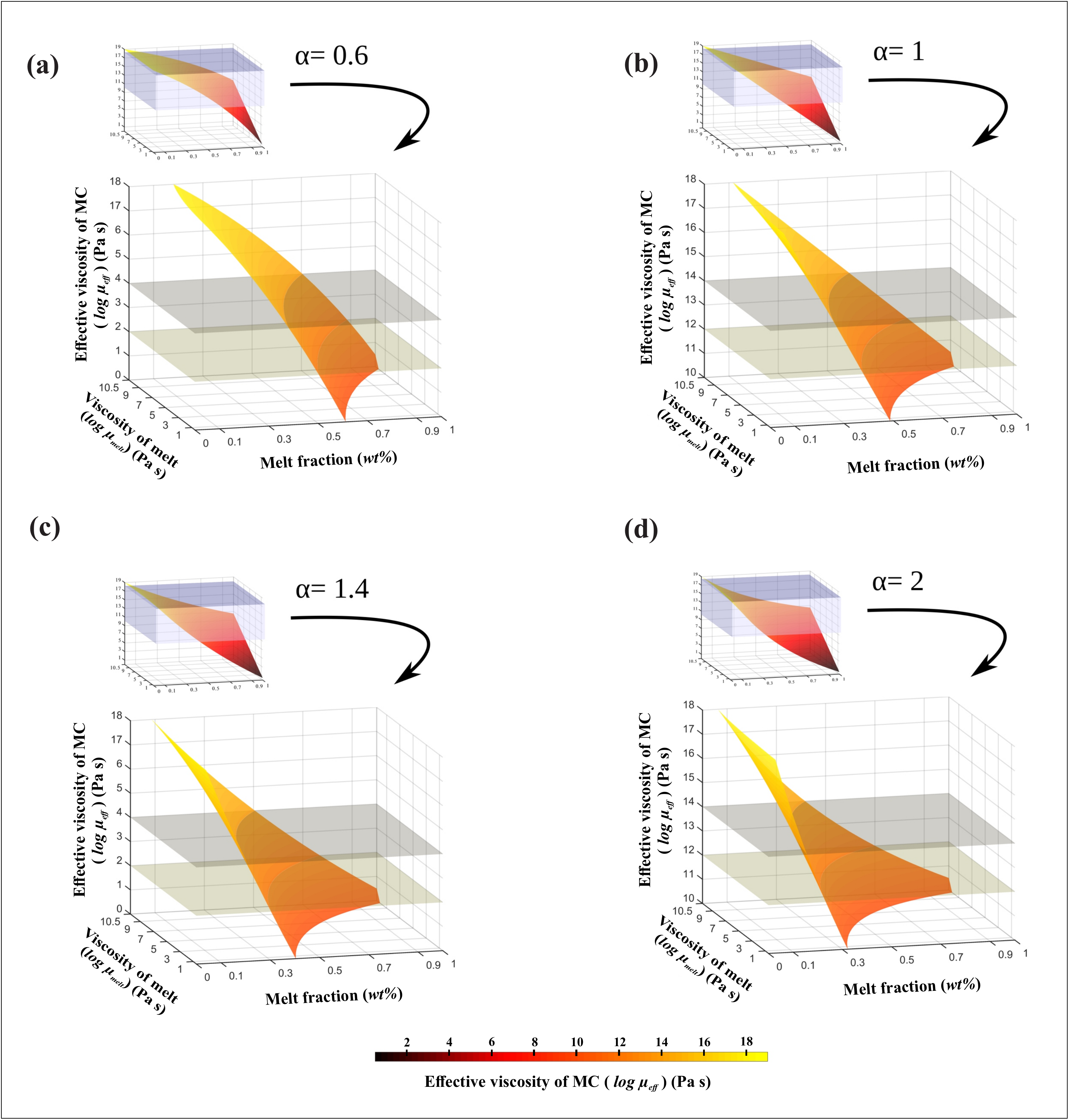}
\caption{Three-dimensional plots of the effective viscosity of MC ($\mu_{eff}$) as a function of
melt suspension viscosity ($\mu_M$) and molar volume fraction ($\phi$), obtained from the two-stage
viscosity calculations for increasing values of the cohesion parameter, $\alpha$. (a) $\alpha$ = 0.6,
characterized by convex surface plot. (b) $\alpha$ = 1 (an ideal situation). (c) and (d) $\alpha$ = 1.4 and 2, respectively. Note the transformation of convex to concave curvature of the surface plot with
increasing $\alpha$. Insets show the full-length plots, highlighting the effective viscosity range of
$10^{10}$ to $10^{18}$ Pa s. Two limiting effective viscosity ($\mu_{eff}$) values, $10^{12}$ Pa s and $10^{14}$ Pa s are shown to constrain the viscosity range in our FSI model to reproduce the spectrum of axial high to flat topography. Host rock viscosity is chosen $10^{19}$ Pa s in these
calculations, emulating mantle.}
\label{Fig2}
\end{figure}

\subsubsection{Melt suspension viscosity}

We introduce a scale of the magmatic process with characteristic lengths, 0.01 to 1 km, as applicable to shallow level magma conduit dimensions (diameter and length) in the sub-ridge region \cite{Head1996}. Lava eruption episodes determine the characteristic time through the conduits. According to the Volcanic Explosivity Index (VEI) \cite{Newhall1982} study, covering more than 75\% of the documented Holocene eruptions, almost 50\% of them record continuous blast duration of less than 6 hrs. Furthermore, 63\% of the eruptions yield eruptive volumes ranging from 0.001 to 0.1 $km^{3}$. Considering the median of time intervals, successive eruptions of a volcano is found to occur in a time-frequency of 13 years \cite{Siebert2015}. We use these data to calculate the strain rates associated with magma flows in shallow conduits. To simplify the calculation, we consider 0.001- 0.1 $km^{3}$ magma undergoing eruption through a cylindrical conduit of 0.01-1 km diameter and on a time duration of 13 years. This wider range of diameters and conduit lengths is chosen because eruptions generally occur through multiple magma conduits of varying lengths, and a cumulative effect of the aforesaid parameters is required in our present analysis. For further simplification, we deal with a representative set of their values, where the magma conduit length ($l_{c}$) and diameter ($d_{c}$) are 1 km and 0.1 km, respectively, and the erupted magma volume ($v_{c}$) is 0.01 $km^{3}$, which passes through the conduit on a time scale ($t_{c}$) of 13 years. This set of values yields an average characteristic strain rate ($\varepsilon^{'}$) of shallow-conduit upwelling of magma, $\varepsilon ^{'} = \left(v_{c} /(0.25\pi{d_{c}}^2l_{c}t_{c})\right) = 10^{-9} s^{-1}$. The characteristic strain rate $\varepsilon ^{'}$ can range from $10^{-6}$ to $10^{-12} s^{-1}$ if the parametric values were varied to cover the entire range discussed above. It is noteworthy that this strain rate scale supports suspended crystals to move passively within the melt phase \cite{CARICCHI2007}.
It is necessary to treat lava eruption on a different time scale (designated as \emph{lava scale}), which represents the duration of a continuous flow event. These events usually take place in a duration of 1 to 12 hours \cite{Siebert2015}. We thus choose an average value of 5 hours to represent the lava scale. Considering the lava conduit length and cumulative diameter as 0.1 km and the erupted magma volume as 0.01 $km^{3}$, we obtain the characteristic strain rate $10^{-3} s^{-1}$ (Table\ref{ Table 1}).\par

The suspension parameters, polymodality, and polydispersity can increase the viscosity of melt suspensions. Now, we calculate the degree of viscosity increase possible in melt suspensions under a sub-crustal environment at MORs. Several studies have provided empirical relations to express the melt viscosity as a function of suspension properties \cite{Einstein1906,Maron1956,Krieger1959,Chong1971,Shapiro1992,Costa2005,Moitra2015}. Consider first the effect of crystal content in melts. Costa \cite{Costa2005} enumerated crystal-free melt viscosity, $\mu_{l}= 10^{5}$ Pa s at a temperature of $800^\circ C$ and a pressure of 300 MPa. The effective viscosity of melts increases with increasing solid volume fraction ($\phi_{s}$) in the suspension. \cite{Marsh1981} suggested that melts erupt as lava with a maximum viscosity, $\mu_{M}= 10^{7} Pa s$ corresponding to $\phi_{(s,c)} = 0.55$, called a critical solid fraction. However, \cite{van1979l} suggested that the critical solid fraction can be further large, $\phi_{(s,c)} = 0.6 \sim 0.7$ at the time of lava eruption, implying crystal-bearing melt viscosity, $\mu_{M}= 10^{8.5}$ Pa s, which means the enhancement of suspension melt viscosity by an order: $I_{1} = 3.5$ \cite{Costa2005,Gonnermann2007}.\par

 Polydispersity ($\delta$) is a measure of the size variation of suspended solid particles in magma. For packing with particle distribution on radii, $P(R)$, the parameter can be expressed as, 
\begin{equation}\label{eq:Poly}
\delta = \frac{\sqrt{\left\langle {\Delta
R}^{2} \right\rangle}}{\left\langle R \right\rangle}
\end{equation}
where ${\delta R } = R-{\left\langle R \right\rangle}$, and the moments of $R$ is defined by $R^{n} > \int R^{n}  P(R)dR$ \cite{Desmond2014}. It is noteworthy that an increase in $\delta$  allows the suspension to increase the   maximum limit of critical solid fraction $\left(\phi_{(s,c)}\right)$. The polydispersity, in turn, multiplies the suspension viscosity. The maximum packing ratio of mono-dispersed spheres accommodates a maximum solid fraction of 0.64, which can increase to 0.75 for suspensions with a polydispersity of 0.65 \cite{BERNAL1960}. Roscoe \cite{Roscoe1952} derived a couple of equations using experimental results \cite{Eilers1941,Ward1950}, showing that various size distribution of rigid spheres influences the viscosity of suspensions less than a uniform size distribution. However, we consider here the theoretical work of Klein \emph{et al.} \cite{Klein2018}, who showed that polydispersity would steeply increase the maximum packing fraction after a threshold limit for monomodal size distributions. This packing effect results in an exponential increase of the suspension viscosity and multiplies its magnitude 40 times. Moreover, the experimental study suggested that an increase in crystal polydispersity might augment volcanic lava viscosity up to 3 orders of magnitude at a higher deformation rate \cite{Moitra2015, Roche2019}. We thus consider the maximum viscosity enhancement in the order, $I_{2} = 3$, corresponding to the polydispersity of crystal-bearing melts.\par

 Strain rate is another factor in our viscosity calculation. Experimental studies suggest Newtonian melt rheology prevails at strain rates lower than $10^{-5} s{-1}$ \cite{Caricchi2011}. But, at higher strain rates, the melts develop shear thinning behaviour \cite{Webb1990}, which  reduces the viscosity by more than 2.5 orders in case of larger solid fraction $\left(\phi_{(s,c)} \sim 0.8 \right)$ \cite{CARICCHI2007}. Considering the strain rates in the order of $10^{-6}$ to $10^{-12}$ $s^{-1}$ on the magma scale and $10^{-3}$ $s^{-1}$ on the lava scale, we choose a maximum viscosity enhancement in the order, $I_{3} = 2.5$, solely due to the decreasing strain rate, leaving out other variables \cite{CARICCHI2007}.\par
 
 To summarize, we use a suspension factor $(I)$, taking into account the cumulative effects of solid crystal fraction ($I_1$), size distribution (polydispersity) ($I_2$), and strain rate ($I_3$), respectively. Considering pure melt viscosity in the order of $10^5$ Pa s, as an example, the suspension viscosity ($\mu_{M}$) can be enhanced to a maximum extent of $10^8.5$ Pa s for a limiting solid fraction (0.6 to 0.7), implying that $I_1$ = 3.5 \cite{Costa2005, Gonnermann2007}.On the other hand, an increase in crystal polydispersity can multiply $\mu_{M}$ by an order of $10^{3}$ at a higher deformation rate \cite{Moitra2015, Roche2019}. We thus consider $I_2$ = 3.  Finally, for the strain rate effects, $\mu_{M}$ can multiply by a factor of $10^{2.5}$ depending on the variation of strain rates in the range $10^{-3}$ to $10^{-12}$ $s^{-1}$, as applicable to the MC in our model. That means, $I_3$ = 2.5. Taking their net effects (i.e., $I_1$ + $I_2$ + $I_3$), we obtain $I = 9$.

\subsubsection{Viscosity of mush complex}

We are now estimating the viscosity ($\mu_{eff}$) of mush complexes (MC), using the theory of mixture rheology within a framework of continuum mechanics \cite{Zhmud2014}. Consider a mixture of host rock ($\mu_{R} = 10^{19}$ Pa s) and melt suspensions ($\mu_{M} = 10^{0.5}- 10^{11.5}$ Pa s), the effective viscosity of MC ($\mu_{eff}$) can be expressed  by the Lederer-Roegiers equation for a two-phase liquid system as,
\begin{equation}\label{eq:LD}
\ln{\mu}_{12} = \frac{x_{1}}{x_{1} + {ax}_{2}}\ln\left( \mu_{1} \right) + \frac{\left( {ax}_{2} \right)}{x_{1} + {ax}_{2}}\ln\left( \mu_{2} \right)
\end{equation}
where $\alpha$ is a constant used to represent the difference in intermolecular cohesive energy between the participating two components, 1 and 2. $x_i$ and $\mu_{i}$ (i = 1, 2) are the mole fraction and the viscosity of $i^{th}$ component in the mixture, respectively. The Lederer-Roegiers equation provides an accurate viscosity calculation of multi-phase fluids with contrasting component viscosities \cite{Zhmud2014}. Equation (\ref{eq:LD}) is close to the Arhenius equation, which can be demonstrated from Roegiers and Zhmud’s \cite{Roegiers2011} approach. Fluidity (inverse of viscosity) of a fluid phase depends on the molar flow activation energy, \(\Delta E\) (a measure of intermolecular cohesion). The Arrhenius relation describes the fluidity in the framework of Eyring’s rate process theory (\cite{Glasstone1941}) as,
\begin{equation}\label{eq:ERT1}
\frac{1}{\mu} = \frac{K}{\widehat{h}}\exp\left( \frac{\mathrm{\Delta}E}{RT} \right)
\end{equation}
which leads to, 
\begin{equation}\label{eq:ERT2}
\ln\mu_{i} = C_{1} + \frac{\Delta E_{i}}{RT}\
\end{equation}
where $C_{1}$ is a constant, $\widehat{h}$ is Planck's Constant, $T$ is absolute temperature, $R$ is the universal gas constant, and $K$ is the ratio of molar volume and Avogadro number. Subscript $i$ refers to the fluid component.\par
For a two-phase liquid system, we consider an additive principle to find the net activation energy of the mixture. According to Eyering’s Rate Process theory of viscosity \cite{Glasstone1941}, the relative motion of one fluid layer over the other demands a molecule to overcome a potential-energy barrier, called flow activation energy per molecule. The total flow activation energy is obtained by taking a product of this quantity with the number of molecules in the system, neglecting any energy dissipation during the molecular transport. Based on this assumption, the total flow activation energy follows,
\begin{equation}\label{eq:AE}
\Delta E_{12} = x_{1}\Delta E_{1} + x_{2}\Delta E_{2}\
\end{equation}
Using equations (\ref{eq:ERT2}) and (\ref{eq:AE}), we arrive at the Arrhenius equation for the binary mixture viscosity,
\begin{equation}\label{eq:AEB}
\ln\mu_{12} = x_{1}\ln\mu_{1} + x_{2}\ln\mu_{2}
\end{equation}
Equation (\ref{eq:AE}) can be generalized with an asymmetric mixing rule (Roegiers and Zhumd 2011) for the flow activation energy:
\begin{equation}\label{eq:AEAs}
\Delta E_{12} = \frac{(1 - \gamma)x_{1}}{(1 - \gamma)x_{1} + \gamma x_{2}}\Delta E_{1} + \frac{\gamma x_{2}}{(1 - \gamma)x_{1} + \gamma x_{2}}\Delta E_{2}
\end{equation}
where $0 < \gamma < 1$. For $\gamma < 0.5$, the contribution of component 1 to the flow activation energy is greater than that of component 2, and vice-versa for $\gamma > 0.5$. Using equations (\ref{eq:ERT2}) and (\ref{eq:AEAs}), we obtain the Roegiers equation (\ref{eq:LD}) by replacing $\alpha = \gamma/(1-\gamma)$ in equation \ref{eq:LD}. $\alpha=1$ implies an equal contribution of flow activation energy by the components, whereas  $\alpha\neq1$ indicates their unequal contributions. For asymmetric two-liquid mixtures, Roegiers and Roegiers \cite{Roegiers1946}, and Roegiers \cite{Roegiers1951} considered $\alpha$ as the ratio of the specific intermolecular attraction energies of the components to derive Equation (\ref{eq:LD}), where $\alpha$ was held constant for an ideal binary system at a particular temperature. The equation, validated experimentally by Roegiers \cite{Roegiers1951}, and later tackled analytically by Zhmud \cite{Zhmud2014} yields $\alpha$ as the ratio $\ln(\mu_{12}/ \mu_{1}) / \ln(\mu_{2}/ \mu_{12})$ for a two-phase system with equal mole fraction of the participating components. In the foregoing analysis we use equation (\ref{eq:LD}) with  $\mu_{1} = \mu_{R}$ and $\mu_{2} = \mu_{M}$, $x_2 = \phi$ (molar volume fraction of melt suspension, and $\mu_{12} = \mu_{eff}$ (MC viscosity).\par 
A set of 3D graphical plots presents the calculated $\mu_{eff}$ as a function of $\phi$ and $\mu_{M}$ for $\alpha$ = 0.6, 1.0, 1.4 and 2 (Figures \ref{Fig2}a-d).  All of them show an inverse relation of the MC viscosity ($\mu_{eff}$) with melt volume fraction ($\phi$) and suspension viscosity ($\mu_{M}$), as widely reported in the literature \cite{Costa2005, Gonnermann2007, Moitra2015}, for the entire range of $\alpha$ values considered in the present calculations. $\mu_{eff}$ is reduced by two orders ($10^{14}$ to $10^{12}$ Pa s) depending on the $\phi$ and $\mu_{M}$ variations. Our model calculations suggest that $\mu_{eff}$ can increase with suspension melt fraction in specific conditions, e.g., suspensions with large volume fractions of crystals, as observed in magmatically robust ridge settings at fast spreading ridges where magmas are extremely enriched with crystals. This model provides the MC viscosity estimates also in opposite environments in slow spreading ridges, characterized by magma poor and low in crystal content, where crystals readily settle down in the course of magma ascent \cite{Lange2013}.

\section{Axial topography: fluid-structure interaction (FSI) modelling }

\subsection{Numerical methods}

\begin{figure}[ptbh]
\includegraphics[width=\textwidth]{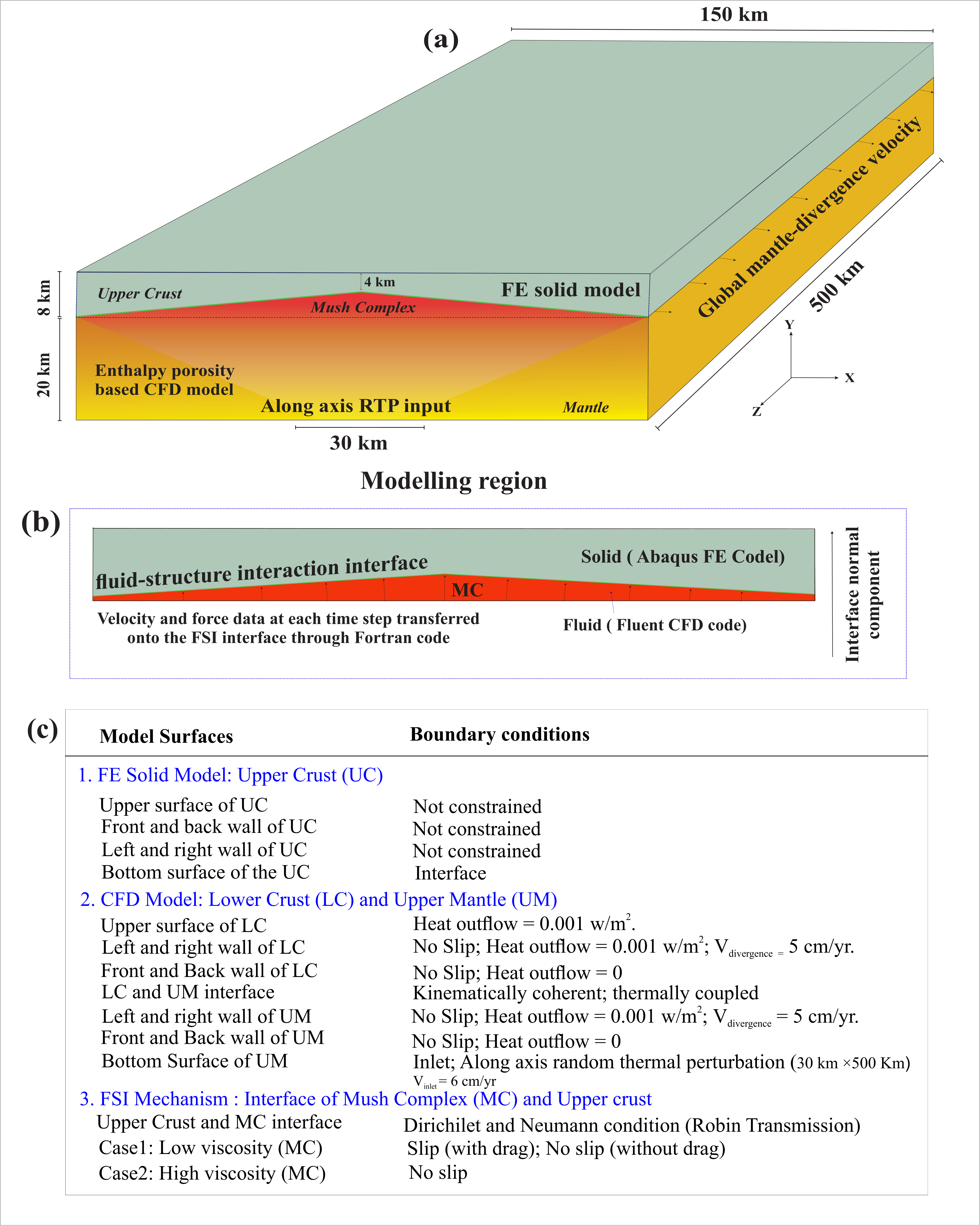}
\caption{(a) Consideration of a three-dimensional fluid-structure interaction (FSI) model for
numerical simulations of ridge-axis topography consisting of fluid subdomain and solid upper
crust. (b) A schematic illustration of Robin-Neumann transmission is used to implement one-
way fluid-structure interaction between mush complex (MC) and the overlying
elastic crust. The CFD model for the fluid regime consists of 2,62,500 nodes, whereas the FE
model for the solid structure consists of 96,635 nodes. The whole numerical calculations
were implemented in a multinode and multiprocessing computer. Each of the FSI coupled FE
simulations took a clock time of 504 hours, preceded by a common CFD simulation, which
took a clock time of 192 hours. (c) All the model boundary conditions (mechanical and
thermal) are summarized in the inset.}
\label{Fig3}
\end{figure}

\begin{figure}[h]
\includegraphics[width=\textwidth]{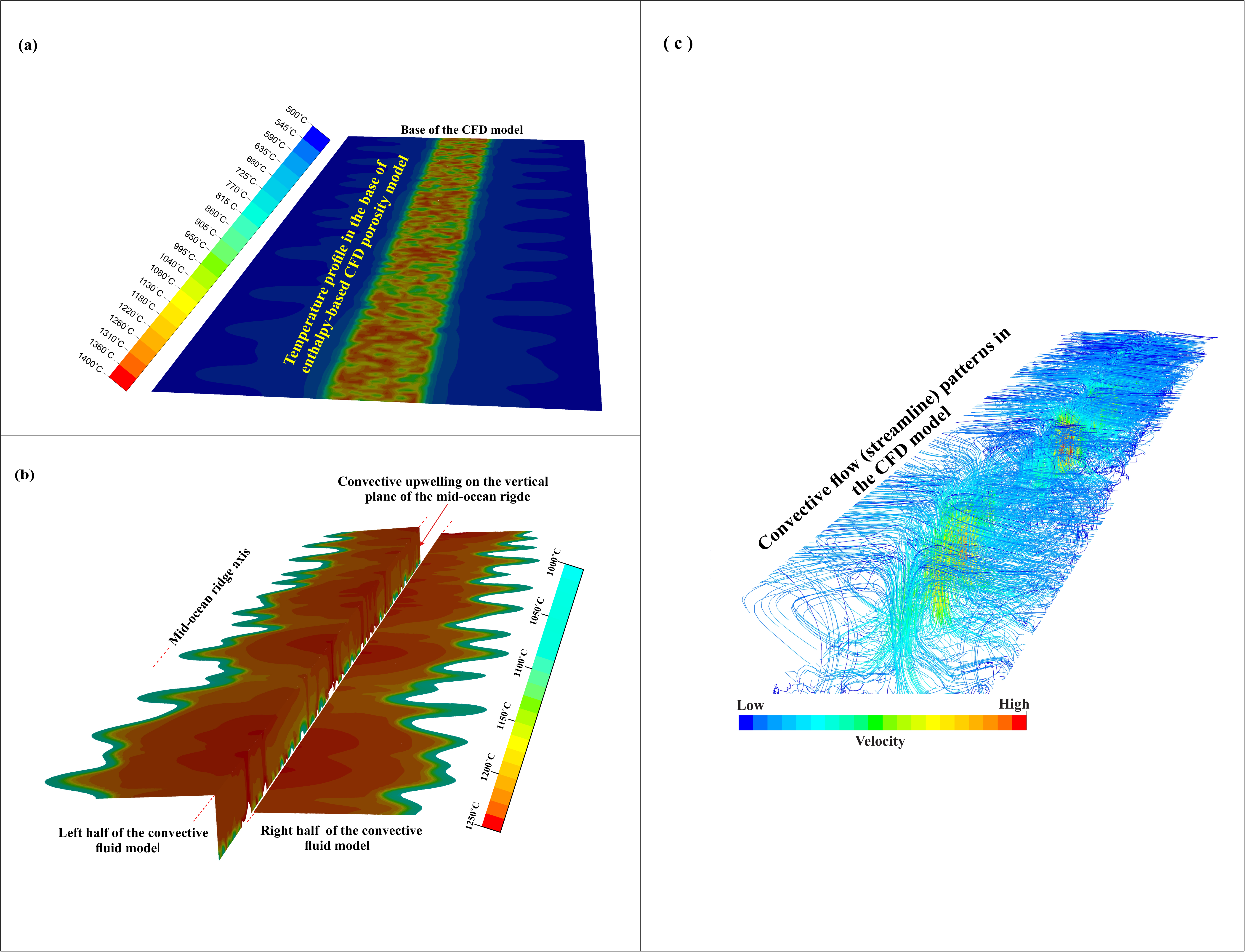}
\caption{(a) Thermal boundary condition imposed at the CFD model base (inlet). Random thermal perturbations (RTPs) are imposed on a 30 Km wide stretch of the model base. (b) Thermal maps of the MC, showing temperature contours on a sub-horizontal plane close to the interface and an along-axis vertical plane, obtained from a CFD model run at 3 Myr. Notice that, 3D convective upwelling is evident from this thermal structure. (c) Streamlines of the 3D convection structures in MC. Both the velocity and the thermal structures show asymmetric 3D convective flows in the model.}
\label{Figure S3}
\end{figure}

This model couples the three-dimensional convective melt upwelling in the melt-bearing mantle part (fluid region) with the overlying elastic layer (oceanic crust) (Figure \ref{Fig3}) in the framework of a Fluid-Structure Interaction (FSI) theory. The fluid sub-problem is tackled using the finite volume computational dynamics code Fluent®, where the CFD model idealizes the mechanical setting as a two-layer system: basal layer (uppermost mantle part), thermomechanically coupled with an overlying high-viscosity layer (i.e., elastic solid crust) (Figure \ref{Fig3}). The model base is subjected to thermal perturbations to simulate thermo-chemical convection with synchronous Darcy’s (porous melt flows) and crystallization (phase transition) \cite{Sarkar2014, Mandal2018}. We then take an average of the three-dimensional velocity data, calculated at the interface above the MC region, and use as the fluid structure interface velocity to set a mechanical (FSI) coupling of the fluid domain with the top elastic crust in the finite element (FE) simulations. This FSI coupling principally aims to reproduce finite deformations in the crustal layer, which otherwise cannot be implemented through the control-volume based fluid simulations used in this study.  It is noteworthy that this combined fluid-solid (CFD-FSI-FE) modelling approach (Figure \ref{Fig3}) geophysically conceptualizes the MC (prismatic sub-axial zone) as a control volume that conserves mass and momentum by a combination of material influx from below and solidification / recycling / eruption, and partly outflux across the top surface \cite{Mandal2018}. The melts in the fluid domain ascend with a complex heterogeneous pattern due to the 3D convection structures (Figure \ref{Figure S3}b), and consequently form ellipsoidal magma pockets with circular plan views, as seen in the FE results (see, Figure \ref{Fig5}a). The mechanical properties of MC are allowed to evolve with the convection in the overall fluid domain, but maintaining both the mass (continuity) and momentum conservations. The theoretical framework of convection in sub-ridge fluid domains is developed on the following conservation equations: continuity, momentum and energy equations, where we introduce a number of source terms: Darcy and buoyancy source terms in the momentum equation, and an enthalpy source term in the energy equation \cite{voller1987,brent1988}. The continuity, momentum and energy equations are finally expressed as follows,
 \begin{equation}\label{eq:Continuity}
 \nabla \cdot v = 0
\end{equation}                              				
\begin{equation}\label{eq:Momentum} 
\rho\frac{\partial}{\partial t}v + \rho v.\nabla v = - \nabla p + \mu_{fd}\nabla^{2}v + S_{g} + S_{D}
\end{equation}
\begin{equation}\label{eq:Energy}
\rho\frac{\partial}{\partial t}{(\rho h)} + \nabla \cdot {(\rho v h)} = \nabla {a\nabla{h}} - S_{h}
\end{equation}       			 
where $p$, $\rho$ and $\mu_{fd}$ denote pressure, density and viscosity of the fluid domain, respectively. $T$, $h$ and $a$ represent temperature, enthalpy, and thermal diffusivity ($a = k/\rho c$, $k$ and $c$ are the thermal conductivity and specific heat, respectively). The fluid velocity, $v$, is chosen to vary linearly with the melt fraction, $\phi$. In this single-phase idealization, the domain viscosity $\mu_{fd}$ is varied as a power-law function of temperature \cite{Sarkar2014,Chen1990}.  In the momentum equation \ref{eq:Momentum} $S_{D}$ regulates the dominance of Darcy (i.e., porous) flow, whereas $S_{g}$ implements the buoyancy factor through Boussinesque approximation. In equation\ref{eq:Energy} acts as an enthalpy factor to incorporate the energy involved in the phase (solid-melt) transformation.

The mathematical expressions of these source terms are,
\begin{equation}\label{eq:SD}
 S_{D} = - C\frac{(1 - \phi)^{2}}{\left( \phi^{3} + \varepsilon \right)}v
\end{equation}
\begin{equation}\label{eq:SG}
 S_{g} = \rho g \theta \Delta {T}
\end{equation}
\begin{equation}\label{eq:Sh}
 S_{h} = \frac{\partial}{\partial t}{(\rho \Delta H)} + \nabla \cdot {(\rho v \Delta H)}
\end{equation}
$C$ and $\varepsilon$ in equation\ref{eq:SD} are constants, whose values are taken as 1e5 and 0.001 respectively, after \cite{Sarkar2014}. In equation\ref{eq:SG}, $\Delta T$ represents temperature fluctuations with respect to the reference temperature, and $\theta$ is the co-efficient of thermal expansion. In equation\ref{eq:Sh}, $\Delta H$ is the mean latent heat content.\par

The fluid subdomain base is subjected to a random thermal perturbation (RTP) condition, which aims to initiate convective flows in the sub-crustal region (\cite{Sarkar2014}). In this random thermal condition partial melting occurs in the domains of high temperatures ($>$solidus), whereas solidification in the domains of low temperatures ($<$solidus) (Figure \ref{Figure S3}a). The initially random thermal state triggers multiple convection cells with three-dimensional asymmetric structures beneath the axis, coupled with porous flows (see Figures \ref{Figure S3}b-c). The porous convection involves in melting and solidification mediated by enthalpy transfer. This thermo-mechanical process leads to a complex along-axis interaction among 3D convection cells in the MC region, resulting in asymmetric subcrustal/lower crustal flow environment (Figure\ref{Figure S3}c). Earlier studies have also reported similar asymmetric nature of porous convection in mid oceanic ridge system \cite{Katz2010}, where the degree of flow asymmetric can intensify by the intervention of synkinematic melting-solidification processes \cite{Sarkar2014}. The asymmetric 3D convection structures in the fluid subdomain eventually results in strong temporal as well as spatial heterogeneities in terms of flow velocity and strain rate in the MC, which in turn give rise to segmented topographic morphologies and axial offsets at MORs, which we elaborate later. In the CFD simulation the 3D convective flows tend to vanish at the top boundary (kinematically coherent interface with the overlying upper-crust), however, remain fully active in the adjacent layers (i.e., MC), allowing the material advection to replenish the MC from below, and maintain a steady state physical condition. The prismatic MC region beneath the ridge axis exhibits the plan views of the 3D convection structures, as described in Figure\ref{Figure S3}b.The mass conservation formulation in the context of axial magma budget calculations has been elaborated in Mandal \emph{et al.} \cite{Mandal2018}.\par

\begin{table}[!h]
\caption{Model Parameters used in thermomechanical model}
\label{Table 2}
\vspace*{4pt}
\begin{tabular}{llll}
\hline
Model &Properties \\
\hline
{\bf FE Model of Upper Crust} &Model dimensions = $4-8~km~\times~150~km~\times~500~km$;  \\
&Density ($\rho_s$) = $2400~kg/m^3$; Elastic modulus ($E$) = $10~GPa$;\\
&Poisson’s ratio ($\nu$) = $0.26$ \\ \\
{\bf FSI Robin-Neumann} &Viscosity ($\mu_{eff}$) = $1 \times 10^{12}~Pa~s - 1 \times 10^{14}~Pa~s$ \\ 
{\bf  transmission} &Density ($\rho_{eff}$) = $2500~kg / m^3$ \\ \\
{\bf CFD model of Upper Mantle} &  $20-24~km~\times~150~km~\times~500~km$;    \\
{\bf and Lower crust} & Viscosity of the upper-mantle (Single Phase Idealization  \\
& Temperature dependent) = $10^{15}~Pa~s$ (Viscosity of mantle \\
& rock = $10^{19}~Pa~s$; Viscosity of melt = $10^3~Pa~s$) \\
& Density of the upper mantle = $2500 kg / m^3$ (Boussinesq); \\
& Thermal Expansion Co-efficient = $5 \times 10^5 /~{}^{\circ}C$; \\
& Thermal diffusivity = $10^{-6}~m^2 /~{}^{\circ}C$; Permeability = $10^{-5}~m^2$; \\
& Specific Heat = $1600~J/kg~^{\circ}C$; Solidus Temperature = $1000~{}^{\circ}C$ \\
& Liquidus Temperature = $1250~{}^{\circ}C$ \\
& Viscosity of lower crust = $10^{21}~Pa~s$; \\
& Density of Lower Crust = $2400~kg / m^3$ \\\hline
\end{tabular}
\vspace*{-4pt}
\end{table}

In FSI (fluid-structure interaction) problems, the solid structures and adjacent fluids interact with each other, where the interaction can be treated with D'Alembert's principle \cite{Belytschko1980} as:

\begin{equation}\label{eq:DALMB}
 \rho {\acute{v}}_i - \sigma_{(ij,j)} + f_{i} = 0
\end{equation}

$f_{i}$ is the body force term, $\rho$ is the density, ${\acute{v}}_i$ is the total time derivative of velocity, and $\sigma_{(ij,j)}$ is the stress tensor gradient. According to the classical mechanics, a fluid-structure interface must obey Dirichlet and Neumann transmission conditions \cite{Hou2012}. In an ideal explicit coupling scheme, the fluid velocity at the fluid-structure interface is determined by the velocity of the structure computed from a structural simulation. On the other hand, the fluid pressure and viscous stress terms modulate the traction on the structure surface at the interface. The velocity transmission onto the fluid system is called the Dirichlet boundary condition. The normal stress transmission onto the structural system is called the Neumann boundary condition. These kinematic and dynamic interface conditions are enforced in the FSI analysis. The Dirichlet-Neumann (D-N) explicit coupling scheme has been successfully worked out for many fluid-structure interaction problems \cite{fernandez2005,Bazilevs2008, Badia2008, Moretti2018}. A combination of Dirichlet and Neumann conditions is often utilized as transmission techniques, instead of pure D-N coupling schemes, depending on the nature of FSI problems \cite{Figueroa2006}.\par
We use a linear combination of D-N transmission conditions (Robin method) in FSI coupling \cite{Nobile2008}. In the present model, the fluid and solid parts have material densities of close values, and similar breadth and width, which is typical in the conforming-mesh partition approach \cite{Badia2008,Hou2012}. But the solid part is much thinner than the fluid domain, allowing us to conceive a simple one-way FSI model. In this approach, the transmission of fluid motifs into the structure at the interface is carried out at each time step and analyzed without a loop-back transmission of structure mechanics onto the fluid and subsequent fluid analysis. The rationale behind this assumption is that movement in the structure part does not create any mechanical feedback effect on the fluid counterpart. This consideration enables us to remodel our problem with a one-way Robin transmission condition in the FSI analysis to simplify the simulations.\par
The interface velocity field obtained from more than 100 time steps (covering $\sim$7 Myr) is coupled with the solid structure to tackle the structural sub-problem separately with the help of an implicit transient finite element code (ABAQUS/CAE®). This velocity field is set at the solid model base to implement specified transmission conditions using a Fortran code developed on Abacus user-subroutines DLOAD and DISP (Figure \ref{Fig3}b). A mathematical framework of this FSI transmission \cite{Badia2008} is given below.\par

The Neumann boundary condition for a structure problem can be written as:
\begin{equation}\label{eq:NB1}
\rho_{s}\delta_{tt}{\hat{w}}^{k + 1} - \nabla{\hat{\tau}}_{s}^{k + 1} = {\hat{f}}_{s}
\hspace{2mm} \text{in}  \hspace{1mm} \Omega_{0}^{s}
\end{equation}
\begin{equation}\label{eq:NB2}
{\tau_{s}}^{k + 1}.n_{s} = - {\tau_{f}}^{k + 1}.n_{f} 
\hspace{2mm} \text {on}  \hspace{1mm} \Sigma^{t}
\end{equation}
where the subscripts $s$ and $f$ denote the solid domain and fluid domain respectively. We use $f(n)$ as an approximation for a time-dependent function $f$ at time level $t(n)$. Backward difference operator $\delta_{t}$ is defined as $\delta_{t}f^{(n + 1)} = \left( f^{(n + 1)} - f^{(n)} \right)/ {\Delta t}$, and $\delta_{tt}(.)= \delta_{t}(\delta_{t}(.))$. $\hat{w}$ denotes displacement in the solid medium with respect to the reference configuration. The superimposed hat symbol indicates the values sought. Superscript $k$ stands for the current iteration; hence ${k+1}$ represents the next iteration. $n_{f}$ is the outward normal to $\Omega_{t}^{f}$ on $\Sigma^{t}$ [fluid-structure interface] and $\eta_{s}= -\eta_{f}$. The solid medium is assumed to be elastic and follows the constitutive relation between Cauchy stress tensor $\tau_{s}$ and deformation gradient tensor $\nabla\hat{w}: F\left( \hat{w} \right) = I + \nabla\hat{w}$. The fluid part is assumed to be homogeneous, Newtonian, and incompressible \cite{Badia2008}. The Cauchy stress tensor is expressed as,
\begin{equation}\label{eq:CST}
\tau_{f}(u,p) = - pI + 2\mu G(u)
\end{equation}
where $p$ is the pressure, $\mu$ is the dynamic viscosity, and,
\begin{equation}\label{eq:ST}
G(u) = \frac{1}{2}\left( \nabla u + (\nabla u)^{T} \right)
\end{equation}
is the strain rate tensor, where $u$ represents the fluid velocity. Using Robin transmission, which is a linear combination of Dirichlet and Neumann components, we obtain a modified set of equations for the structure,
\begin{equation}\label{eq:NB3}
\frac{\beta_{s}}{\Delta t}w^{(K + 1)} + {\tau_{s}}^{(k + 1)}.n_{s} = \frac{\beta_{s}}{\Delta t}w^{(n)} + \beta_{s}u^{(k + 1)} - {\tau_{f}}^{(k + 1)}.n_{f}
\hspace{2mm} \text {on} \hspace{1mm} \Sigma
\end{equation}
where $\beta_s$ has a suitable positive and bounded value that determines the contribution of the Dirichlet component in the Robin transmission \cite{Badia2008}. The bounded positive value of $\beta_s$ is set at 0.025, which is held constant spatially and temporally in a single simulation run, as well as in all the simulations. We assessed impact of this constant on the model calculations, and found little variations on its increasing or decreasing values. The nominal Dirichlet term acts as a perturbation to the crustal base to localize short-wavelength undulations, whereas the Neumann boundary condition gives rise to first-order, long-wave vertical undulations, and localizes 3D deformations in the elastic crust. In order to obtain both long- and short-wavelength axial topography, we impose a linear combination of the two types of interactions at the interface \cite{Figueroa2006, Badia2008}. The Dirichlet terms utilize the velocity of fluid domain at interface, described by a 3D array with three velocity components on one dimension, 18750 spatial points on the second dimension and 111 temporal points (covering 7 Myr) on the third dimension. For the Neumann term, the normal component of a strain-rate tensor at the fluid-solid interface in equations (\ref{eq:CST}) and (\ref{eq:ST}) is obtained from the vertical component of the fluid velocity vector, corresponding to a two dimensional array of spatiotemporal points, considering a very small slope ($3^{\circ}$) of the interface. Our estimate yields a strain-rate median of $10^{-5} s^{-1}$ at the axis, with upper and lower limits, $10^{-3} s^{-1}$ and $10^{-11} s^{-1}$ on the interface, depending on the vertical component of the velocity. The calculated strain rates cover values for the entire lava/magma movement range.\par

Equation\ref{eq:NB3} signifies the introduction of the Dirichlet terms in the Neumann boundary condition equation\ref{eq:NB2} at fluid-structure interface. The Dirichlet term indicates the fractional transfer of fluid kinematics at the interface to the adjoining structure, while the Neumann term indicates the force transfer at normal direction.  Thus the viscous thrust term (Neumann term) almost solely controls the first-order axial topography, whereas the localized short-wavelength variation of reliefs is primarily triggered at crustal base by the nominal contribution of the Dirichlet term (factored by $\beta_{s}= 0.025$) of the Robin transmission of FSI. The rest is done through the material behaviour of the overlying crust. This FSI analysis explains the physical mechanism why a particular effective viscosity at MC would generate a specific topography at MORs (Figure\ref{Figure S4}; Figure\ref{Fig5}). Although the mantle and the overlying crust are primarily considered kinematically coherent, a relative motion between them can develop if the viscosity of the sub-crustal / lower crustal MC is comparatively low. Such kinematic state in turn produces shear stresses at the interface. This viscous drag is treated as an independent variable because it originates from the velocity difference (slip condition) between the moving plate and the underlying horizontal flow in the MC. We include the drag factor selectively in low-viscosity models ($\mu_{eff} < 7.5x10^{12} $ Pa s), where the shear drag is given by,
 
  \begin{equation}\label{eq:Shear}
 \tau_{x} = 2\mu_{eff}{\dot{e}}_x
 \end{equation} 

where ${\dot{e}}_x$ is the horizontal shear rate away from the ridge axis. Equation \ref{eq:Shear} comes directly from Newtonian fluid's interaction with the wall as a shear stress term \cite{Kundu2015}, applied in an area bounded by adjacent nodes. We calculated the horizontal shear rate, considering an average value of the velocities measured at a point on the interface. As the Robin equation does not contain the horizontal viscous drag term \cite{Badia2008}, we introduce this term and update the equation to investigate the additional effects of this stress factor on the axial topography.\par

The pressure term in the Cauchy stress tensor (\ref{eq:CST}) is written as
\begin{equation}\label{eq:Pressure}
p = -\left( \rho gh + \frac{1}{2}\rho u^{2} \right)  
\end{equation}
where g stands for gravity, $h$ is the depth of spatial points at the interface. We use the fluid domain density (equation\ref{eq:Pressure}) in the pressure term of the Cauchy stress tensor (equation\ref{eq:CST}), which is kept constant in the fluid simulation as the thermal expansion is used only in the source term (equations \ref{eq:SG} and \ref{eq:Sh} ).But we find the fluctuation of density has little effect on force transfer to overlying upper crust (elaborated in the concluding paragraph of model results). In summary, We thus take the kinematic output from the CFD calculations and couple with the solid crust for the FSI, excluding any material attributes. The fluid-solid interaction is then defined fully by the mechanical term- effective viscosity ($\mu_{eff})$, calculated independently as presented in the preceding section.\par

The top crustal layer is treated as a 3D elastic solid \cite{Olive2020} with orthotropic engineering moduli in a finite element (FE) model with realistic dimensions (Figure \ref{Fig3}a ). To reveal the exclusive effects of melt-bearing sub-crustal viscosity on the ridge axis topography, we chose not to impose any tectonic and pre-existing boundary conditions on the model lithosphere, and also excluded any auxiliary pull forces, prefixed weak zones, or any brittle fracture zone in the crust \cite{Buck2005}. In addition, we simplify the model by treating the crust as a single mechanical layer, discretized into several layers of one material \cite{Gudmundsson2004}, as this FSI model aims to show the probable effects of sub-crustal mush complexes (MC) on the first-order axial topography of MORs (Figure \ref{Fig1}a). The detailed material properties used for the two model components: FE solid model of crust and the CFD model of uppermost mantle, and their Fluid-Structure coupling, are presented in Table\ref{Table 2}. To obtain the sole effects of MC, we excluded global spreading in the model crust (also discussed in the model limitation). However, an across-axis viscous shear drag to the overlying elastic layer is incorporated in some simulations separately (discuss in preceeding paragraph). The detailed boundary conditions used in the modelling have been provided in the Figure \ref{Fig3}c.\par 

\subsection{Model results}

\begin{figure}[H]
\includegraphics[width=\textwidth]{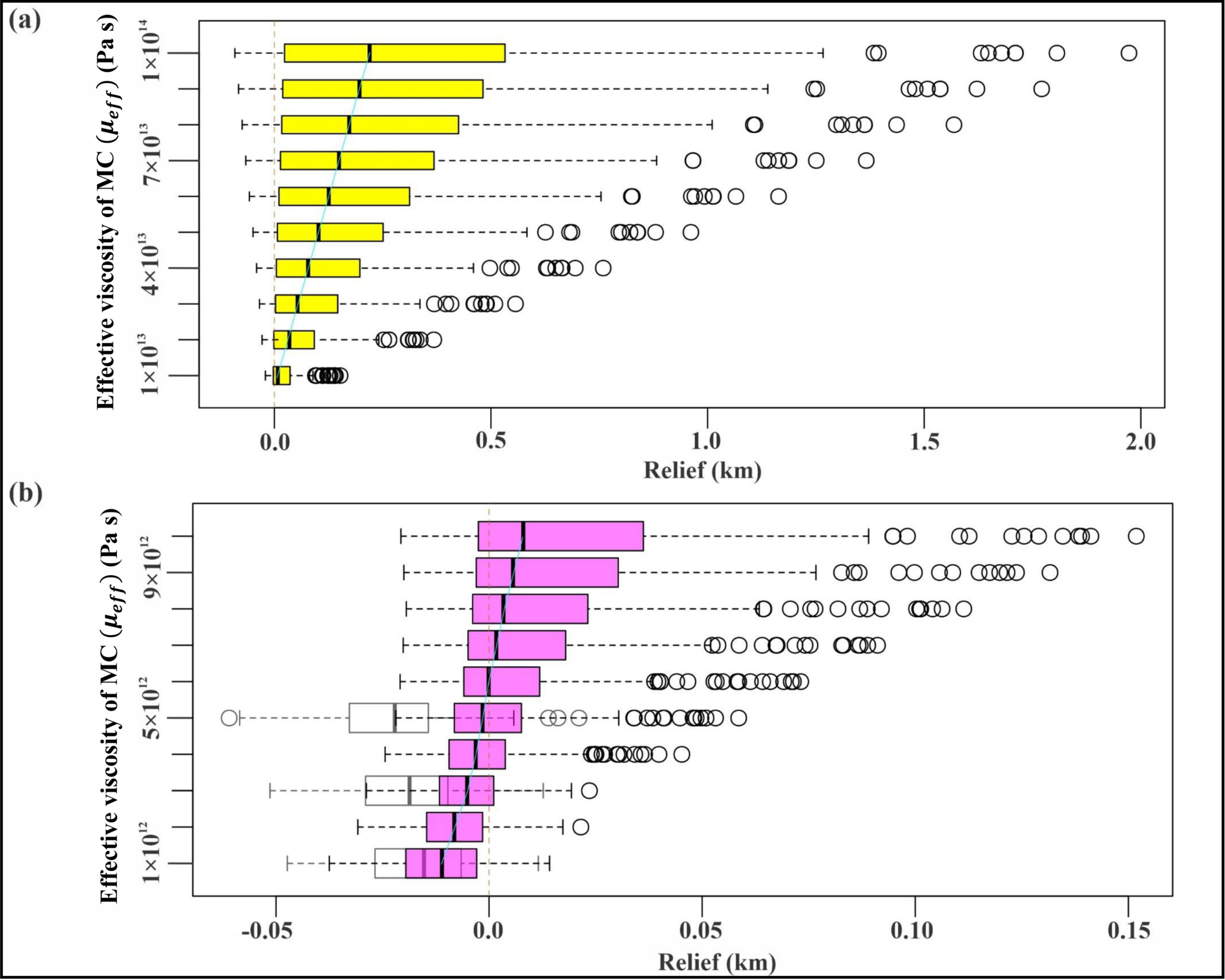}
\caption{Box plots of the vertical displacement at the top layer nodes in the evolution of ridge axis in FE models. The box plots show the median of axial reliefs at nodes on the axis as a thick black line. The coloured box is bounded by $25^{th}$ and $75^{th}$ percentile reliefs, named first quartile ($Q_{1}$) and third quartile ($Q_3$). The box length defines the interquartile range ($IQR$) of data. The dotted line or the whisker delineates the ``maximum";and ``minimum"; reliefs, represented by ($Q_{3} + 1.5\times IQR$) and ($Q_{1 }-1.5\times IQR $), respectively. Empty circles mark some data points, called outliers, beyond the limiting range. (a) Calculated plots of axial relief data obtained from FSI models run with $\mu{eff} = 10^{13}$ Pa s to $10^{14}$ Pa s. (b) A similar plot of the axial relief, but for a lower viscosity range: $10^{12}$ Pa s to $10^{13}$ Pa s. It is noteworthy that the viscosity versus axial relief relation becomes non-linear at $\mu_{eff} = 7.5\times10^{12}$ Pa s. This nonlinearity indicates the weakening of normal viscous force component in the Robin transmission condition. We introduced an across-axis drag force component in the transmission condition for low-viscosity simulations (indicated by white box plots). The box plots in (4b) show a maximum relief of 1.27 km ($\mu_{eff} = 10^{14}$ Pa s) and a minimum relief of -0.06 km ($\mu_{eff} = 5\times10^{12}$ Pa s + drag)}
\label{Fig4}
\end{figure}

We investigated the mode of axial topographic development in a series of FSI model simulations run with varying $\mu_{eff}$ (mush complex viscosity) in the Cauchy stress term of the Robin equation (\ref{eq:NB3}). $\mu_{eff}$ was chosen to vary in the range $10^{12}$ to $10^{14}$ Pa s, as discussed in the preceding section, and this viscosity range gives rise to a broad spectrum of axial reliefs (-0.06 km to 1.27 km at 7 Myr, see box-plot for minima and maxima in Figures \ref{Fig4}a and \ref{Fig4}b), as reported from non-rifted natural ridges. The post-processing of the model results shows that the linear relation between µeff and the median relief breaks as µeff is reduced to a threshold value of 7.5x$10^{12}$ Pa s (Figure \ref{Fig4}b).\par

\begin{figure}[h]
\includegraphics[width=\textwidth]{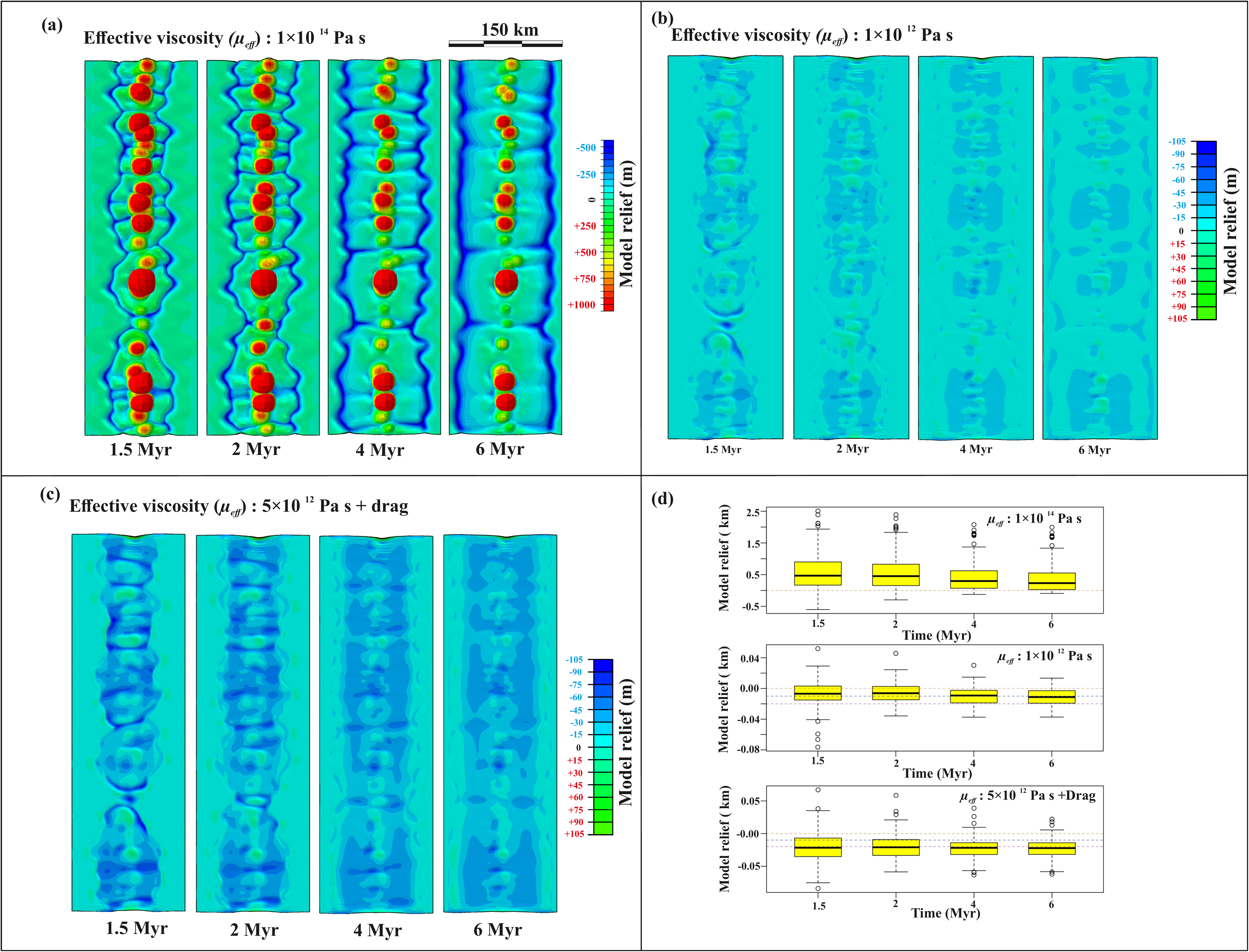}
\caption{Vertical elevation maps showing contrasting MOR topographic patterns in FE
models for varying MC viscosities: (a) high ($\mu_{eff} = 10^{14}$ Pa s) (b) low ($\mu_{eff} = 10^{12}$ Pa s) and (c) low ($\mu_{eff} = 5\times10^{12}$ Pa s), coupled with across-axis drag at the lithospheric base. Model run time: 1.5 Myr, 2 Myr, 4 Myr, and 6 Myr. (d) Box plots of the topographic reliefs calculated at the nodes of the evolved axes are shown in the three panels: (a) to (c). It is noteworthy that the median values change with time.}
\label{Fig5}
\end{figure}

\begin{figure}[H]
\includegraphics[width=\textwidth]{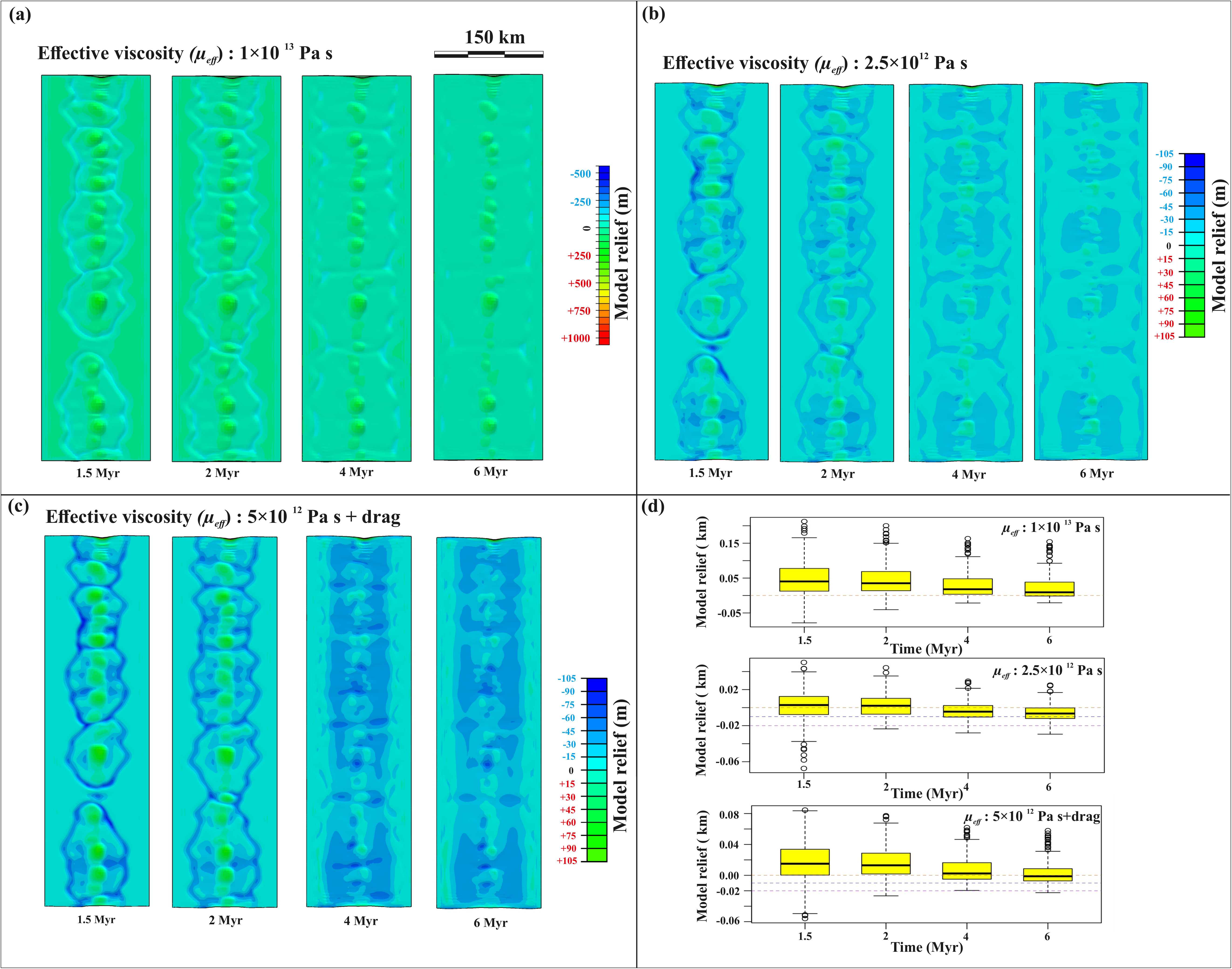}
\caption{Topographic elevation maps of FE model of MOR run with varying MC viscosity ($\mu_{eff}$): (a) moderate ($\mu_{eff} = 10^{13}$ Pa s), (b) low ($\mu_{eff} = 2.5\times 10^{12}$ Pa s)  and (c) low ($\mu_{eff} = 5\times 10^{12}$ Pa s + drag ). Model run time steps: 1.5 Myr, 2 Myr, 4 Myr, and 6 Myr. (d) Box-plots of the reliefs at the nodes of the evolved axes in the three FE models: (a) – (c).}
\label{Figure S4}
\end{figure}

We first present two 3D numerical simulations to show the vertical axial displacement as a function of $\mu_{eff}$ (Figures \ref{Fig5} and also see Figure \ref{Figure S4}). The simulation run with $\mu_{eff}$ = $10^{13}$ Pa s produces a prominent linear zone of upward vertical displacement (axial high) with a maximum elevation of nearly 0.1 km on a time scale of 6 Myr (model run time) (Figure \ref{Figure S4}a). The topographic elevation becomes more than 1 km in the model with $\mu_{eff} = 10^{14}$ Pa s (Figure \ref{Fig5}a). The model results clearly suggest a positive relation of the axial height with $\mu_{eff}$. A time-series analysis (1.5 Myr to 6 Myr) reveals a characteristic temporal variation of the vertical displacement at the ridge axes, especially in their central regions (15 km on either side) (Figure \ref{Fig5} and Figure \ref{Figure S4}). The vertical uplift progressively weakens but remains active in the entire model run-time of 7 Myr (Figure \ref{Fig5}). Both the models produce persistent along-axis linear zones of vertical uplift (axial high) in the initial stages (t = 1.5 Myr). However, the uplift pattern is strongly heterogeneous in the axial direction (Figure \ref{Fig6}a-b), resulting in topographic segmentation of the axial highs, as widely reported from bathymetric studies (\cite{Carbotte2016} and references therein).\par

The median value of axial elevations ($W_{AR}$) at a given model run time (e.g., t = 1.5 Myr) increases consistently with $\mu_{eff}$, resulting in a transition from flat to axial high topography. For example, $W_{AR}$ = -1 m (i.e. almost flat) for $\mu_{eff} = 10^{12}$ Pa s (Figure \ref{Fig5}d), which increases to 40 m at $\mu_{eff} = 10^{13}$ Pa s (Figure \ref{Figure S4}d), and to 467 m (i.e., axial high) when $\mu_{eff} = 10^{14}$ Pa s (Figure \ref{Fig5}d). The axial highs are flanked by a pair of ridge parallel belts of downward vertical movement, forming narrow troughs at a distance of 40 to 60 km from the ridge axis (Figure \ref{Fig5}a and \ref{Fig6}a). The magnitude of average negative relief at the off-axis troughs ($W_{OR}$, median of reliefs at the nodes along axis-parallel troughs) also increases with increasing  $\mu_{eff}$; for example, at 6 Myr, $W_{OR} = - 30 $ m for $\mu_{eff} = 10^{13}$ Pa s, (Figures \ref{Figure S4}a and \ref{Figure S4}d) whereas $W_{OR} = - 350 $ m when $\mu_{eff} = 10^{14}$ Pa s (Figures \ref{Fig5}a and \ref{Fig5}d). A positive relation of $W_{OR}$ with the axial elevation ($W_{AR}$, median of reliefs at axis nodes) suggests a correspondence between the axial-high loading and downward flexural deformation of the elastic crust to produce ridge-parallel depression zones. Model simulations with lower $\mu_{eff}$ (1 - 2.5 $\times$ $10^{12}$ Pa s) show a dramatic change in the evolution of ridge-axis topography. The ridge axis develops a series of 60 to 80 km long discrete depressions ($W_{AR}$  =  - 11 m to -6 m at 6 Myr, Figures \ref{Fig5}d and \ref{Figure S4}d) on a wavelength of 110 to 130 km along the axis, but without any axial highs (Figures \ref{Fig5}b and \ref{Figure S4}b). The simulations with $\mu_{eff}$ varying in the range $10^{12}$ to $10^{14}$ Pa s, thus indicate that the viscosity of sub-crustal mush complex zones critically determines the evolution of flat versus axial high topography ($W_{AR}> 0$) in MORs.  The axial high topography is possible to develop only when the viscosity of sub-crustal mush complexes exceeds a threshold value ($\sim 6 \times 10^{12}$ Pa s), as demonstrated in Figure \ref{Fig4}b, showing crossovers from slightly negative or flat to positive relief at most of the points on the surface.\par

We independently investigated the additional effects of magmatic drag forces on the growth of axial highs. As this factor becomes more effective in the case of a low-viscosity MC condition, we present here two simulations run with low $\mu_{eff}$ (= 5x$10{12}$ Pa s), one with and the other without basal drag factor. The drag-free model (Figures \ref{Fig5}a-b and \ref{Figure S4}a-b) produces length-wise persistent axial highs of moderate elevations ($W_{AR}$ = 13 m at 2 Myr), flanked by low-amplitude ($W_{OR}$ = -50 m) ridge-parallel depressions (Figures \ref{Fig5}d and \ref{Figure S4}d). However, the axial high progressively reduces its average elevation, forming an almost flat topography ($W_{AR}$ $\sim$ 0 m, Figures \ref{Fig5}d and\ref{Figure S4}d) at 6 Myr. The off-axis depressions reduce their negative relative relief to flat ($W_{OR}$ = -25 m). The simulation with basal drag produces axial topography, dominated by a series of depressions, leaving sporadic small highs but not forming any persistent linear topographic high (Figures \ref{Fig5}c and \ref{Figure S4}c). The axial depressions ($W_{AR}$ = -22 m at 1.5 Myr, minima = - 75 m) hardly change their negative relief on a run time of 6 Myr ($W_{AR}$ = - 22 m, minima = - 58 m, Figures \ref{Fig5}d and \ref{Figure S4}d) and form a weak depression, flanked by a flat topographic belt ($W_{OR}$ $\sim$ 0) at a distance of 60 km from the ridge axis (Figures \ref{Fig5}c-d; \ref{Figure S4}c-d)\par

\begin{figure}[ptbh]
\includegraphics[width=\textwidth]{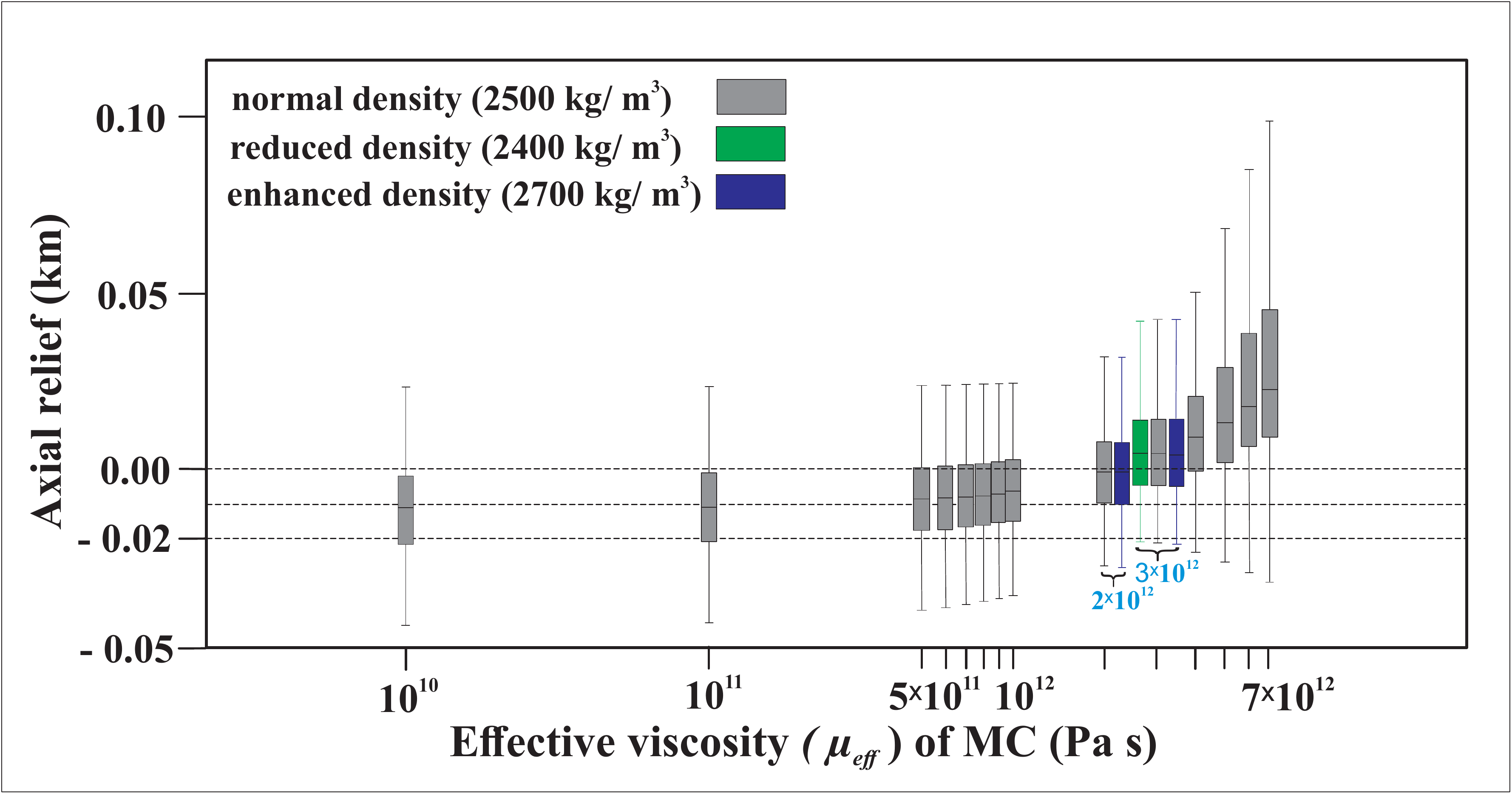}
\caption{Semi-log graphical plots (model run time 3 Myr) of the axial relief as a function of the effective viscosity ($\mu_{eff}$) of MC. Note that, little or no fluctuations occur as the MC viscosity decrease down to a value $\leq 1\times 10^{12}$ Pa s. Also, the differential topographic variations are not sensitive to the MC density at low ($\mu_{eff}$).}
\label{Figure S5}
\end{figure}

We examined the effect of densities of fluid subdomain and MC on the MOR topography. The density of the MC region is varied as a function of the temperature distribution using the thermal expansion coefficient. The extent of density variation ($\sim$ 100 $kg/m^{3}$) is used in the fluid-structure interaction process. Numerical simulations for this variation show little or no effect on ridge topography at low effective viscosities (for example, at $3\times10^{12}$ Pa s, see Figure\ref{Figure S5}). Also, we ran simulations with a sufficiently high density (upto 2700 $kg/m^{3}$) and found very little difference in topography in case of lower effective viscosities (for example at $2\times10^{12}$ Pa s and at $3\times10^{12}$ Pa s) (Figure\ref{Figure S5}). The observed results can be explained by considering the relative magnitude between the pressure term (equation\ref{eq:Pressure}) and the viscous stress term in equation\ref{eq:CST} (main text). The viscous stress term that varies with the strain rate largely controls the morphological undulations, when the pressure terms in Cauchy stress tensor (equation\ref{eq:CST}) remain almost unaffected. The pressure created by flow velocities at the base is $ 0.5 \times \text{density} \times \text{velocity}^2$, where the magnitude of velocity is extremely low, as calculated from the strain-rate range. The dynamic pressure part, involving square of the velocity term, is thus negligible small, as compared to the static pressure ($ \text{density} \times \text{gravity} \times \text{depth}$).
Again, the effect of static pressure becomes relatively weak in case of high viscosity conditions. For example, for a MC viscosity of $10^{13}$ Pa s the calculated viscous stress (equation\ref{eq:Pressure}) is in the order of hundreds of MPa at the interface for an average strain rate of $10^{-5} s^{-1}$, whereas the static pressure is in the order of tens of MPa at the interface, implying that the viscosity will dominantly control the process of topography building in the overlying solid crust. For large effective viscosity of the MC ($>10^{12}$ Pa s), crustal deformations at MORs are thus attributed to the rheological conditions of the subcrustal magmas, rather than the buoyancy conditions in the MC. (Figure \ref{Figure S5}). Thus, the axial highs in our models are not a manifestation of the density structure in the MOR system. This factor only influences the magnitude of flat axial topography under low-viscosity conditions in the MC.

To summarize, the 3D views of a high ($\mu_{eff} = 10^{14}$ Pa s) and a low-viscosity ($\mu_{eff} = 10^{12}$ Pa s) model reveal a spectacular difference in their stable ridge topography produced on a run time of 7 Myr (Figures \ref{Fig6}a-b), which broadly agree with those observed in nature. A time-series analysis of the across-axis profiles of model topography shows that the off-axis troughs continuously migrate away from the ridge axis, leaving a flat region between the axial high and them (Figures \ref{Fig6}a-b). The FSI model explains the mechanics of MOR topographic modulation by $\mu_{eff}$. The Cauchy stress term in the Neumann condition for the FSI consists of two terms – a) hydrostatic pressure b) viscous stress (\ref{eq:CST}). The latter is significantly higher than that the density controlled buoyancy pressure (i.e., the first term). However, the two dynamic terms turn to be in similar orders when the MC viscosity becomes low. For $\mu_{eff} < 10^{12}$ Pa s, the axial topography no longer varies with viscosity; it is the static pressure term (equation\ref{eq:CST}) that takes the control in producing a flat topography (Figure\ref{Figure S5}).\par

\section{Discussions}

\subsection{Effects of sub-crustal melt accumulation}

Using a three-dimensional graphical plot (Figure\ref{Fig2}b), we have shown the effective viscosity ($\mu_{eff}$) of MC as a function of the suspension viscosity ($\mu_{M}$) and the volume fraction of crystal-bearing melts ($\phi$) in the system. An increment of $\mu_{M}$ by an order of 7 (2 to 9), accompanied by an increase of $\phi$ from $\sim 40\%$ to $50\%$, i.e., pure melt fraction anywhere between 8\% and 30\%, would eventually increase $\mu_{eff}$ from $10^{12}$ to $10^{14}$ Pa s (Figure \ref{Fig7}a). This inverse relation of $\mu_{eff}$ with $\phi$ resolves the apparently contradictory observations, axial highs in the magma-rich EPR ridges (\cite{key2013}), and flat ridge topography in the magma-poor MAR. The same explanation applies to the topographic transition, high to flat in SEIR at $103^{\circ}35'$E, where both sides are somehow rifted \cite{Carbotte2016}.

\begin{figure}[H]
\includegraphics[width=\textwidth]{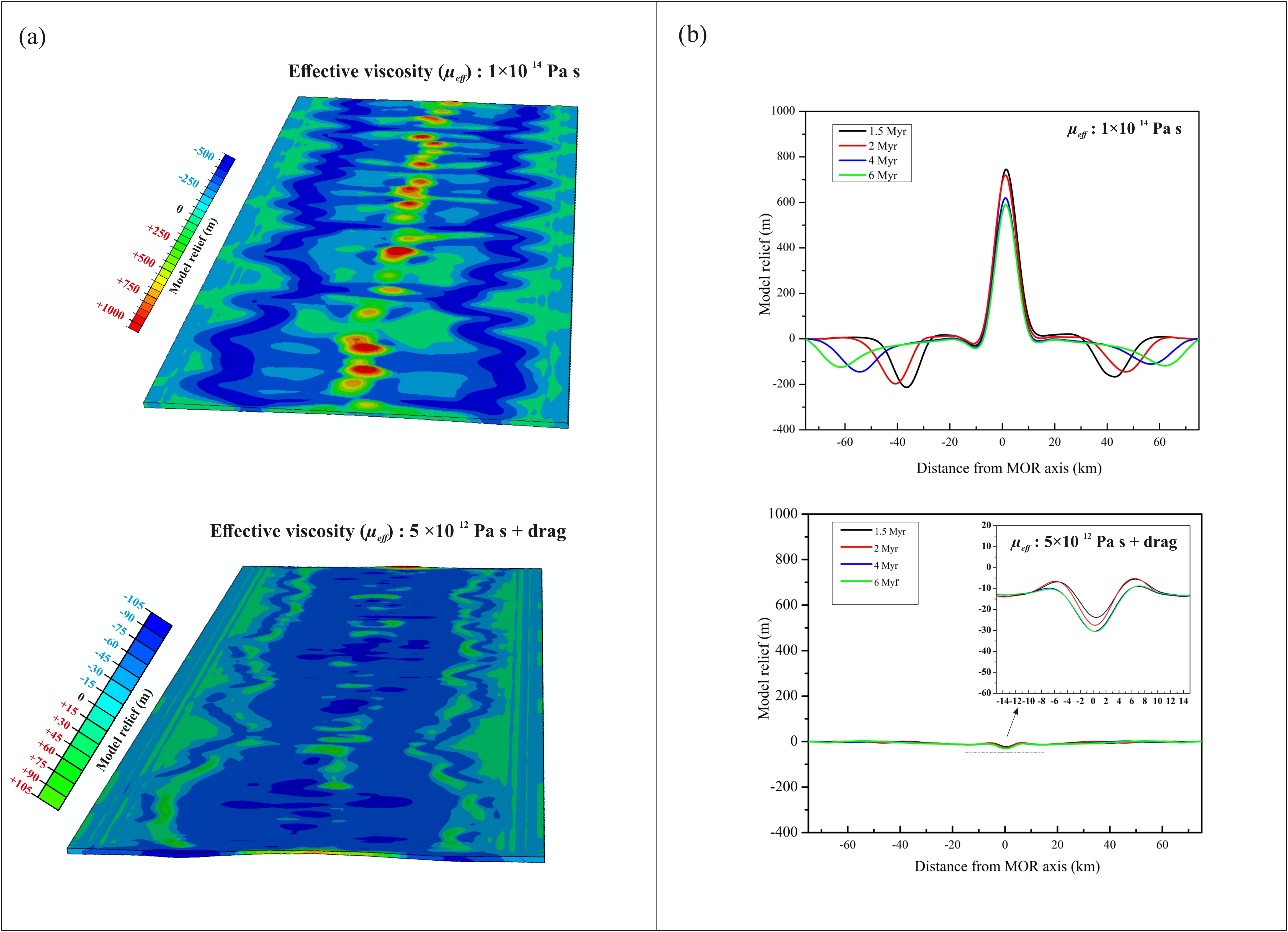}
\caption{(a) 3D views of the axis topography in models with $\mu_{eff} = 10^{14}$ Pa s (upper panel) and $10^{12}$ Pa s (lower panel). Model run time: 7 Myr. The low-viscosity model was run with basal drag force. (b) A time-series analysis of the first-order across-axis topographic profiles from the high- and low-viscosity simulation runs. The inset shows a magnified view of the axial negative relief in the lower panel.}
\label{Fig6}
\end{figure}

\begin{figure}[h]
\includegraphics[width=\textwidth]{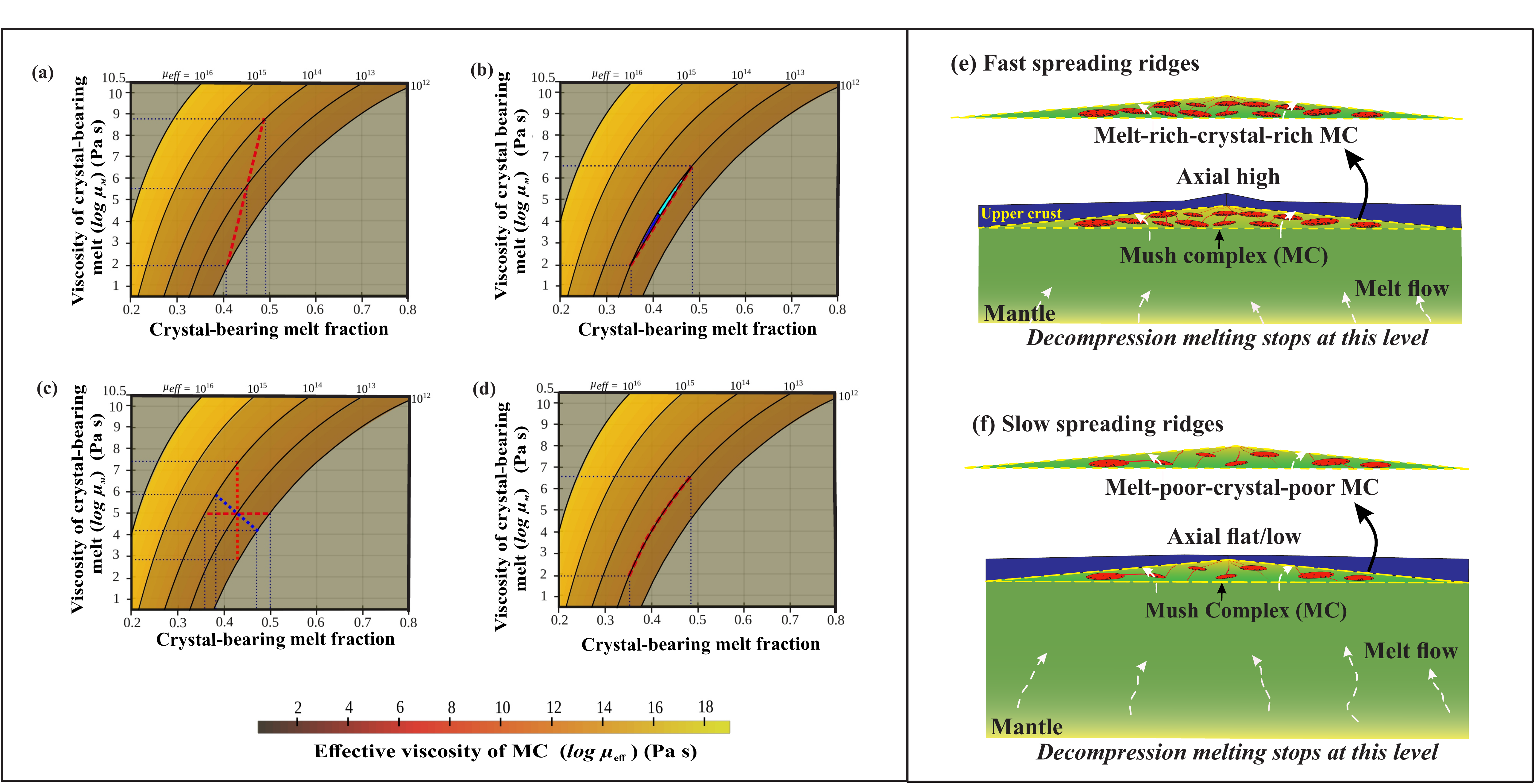}
\caption{Projection of the 3D plots for MC viscosity (presented in Figure \ref{Fig2} 2) on a 2D frame defined by volume percentage ($\phi$) and viscosity ($\mu_{M}$) of crystal-bearing melt (i.e., melt-
suspension). (a) A specific regression where a large increase in $\mu_{M}$ (an order of $10^7$ Pa s) with a moderate rise ($\sim 10\%$) in $\phi$ enhances the effective viscosity ($\mu_{eff}$) of MC by up to two orders. (b) A linear regression of $\mu_{M}$ with $\phi$ involving a limited change in $\mu_{eff}$, where a zone of a small decrease is followed by a zone of little increase (shaded with different colours),
ultimately leading to the same $\mu_{eff}$ under a specific combination of the $\mu_{M}$ and $\phi$ variation. It is to be noted that the red dotted line defines a critical slope line; a regression below this line would result in a lowering of $\mu_{eff}$, whereas above it would increase $\mu_{eff}$. (c) A regression (blue dotted line) of a smaller change in $\mu_{M}$ ($\sim 10^{1}$ Pa s) as well as a lower change ($\sim 5\%$) in $\phi$ showing a modification of $\mu_{eff}$ by up to 1 order of magnitude. Note that a similar order of change in $\mu_{eff}$ is possible when one of the two parameters: $\phi$ or $\mu_{M}$ varies, keeping the other constant, as indicated by red dotted lines. (d) A steady-state condition of $\mu_{eff}$ in complex with varying $\phi$ and $\mu_{M}$ along a particular regression (dashed red line). (e) and (f) Cartoon diagrams of the sub-crustal phenomena at typical fast- and slow-spreading ridges, showing the possibility of higher and lower effective viscosities in a melt-rich-crystal-rich and a melt-
poor-crystal-poor conditions, respectively.}
\label{Fig7}
\end{figure}

 It is noteworthy that the inverse $\mu_{eff}$ -- $\phi$ relation occurs below a threshold slope of the $\mu_{M}$ versus $\phi$ curve, as demonstrated in Figure \ref{Fig7}b. The threshold regression line shows that increasing $\phi$ initially reduces $\mu_{eff}$, followed by a compensatory rise, ultimately attaining the same $\mu_{eff}$ value. Under a threshold condition, across-axis asymmetric sub-ridge melt distributions beneath MORs can thus hardly break their axial topographic symmetry \cite{Evans1999}. Figure \ref{Fig7}c shows different possible paths of $\mu_{eff}$ variations with $\mu_{M}$ and $\phi$. $\mu_{M}$ and $\phi$ (blue dotted line) can locally fluctuate in ridge settings due to some variations in the strain rate. This fluctuation results in an unsteady state of $\mu_{eff}$, ultimately leading to a local instability in the axial topography. In specific cases, $\mu_{eff}$ can remain steady over a broad range of non-linear $\mu_{M}$ -- $\phi$ regression, as shown in Figure \ref{Fig7}d. Such a sub-crustal condition is possibly required for the long-timescale relative stability of axial morphologies, as reported from many MORs \cite{Rowley2016}.\par
 
The viscosity model also yields $\mu_{M}$ -- $\phi$ relations that support contrasting observations from fast and slow spreading ridges; axial high topography in fast ridges with high melt percentages, whereas axial flat topography in slower ridges with low melt contents. We now provide simple numerical estimates to discuss the MC viscosity as a function of crystal-bearing melt viscosity ($\mu_{M}$) and molar volume percentage ($\phi$) of melt suspension using the mixture rheology curve in Figure \ref{Fig7}. For a given value of $\mu_{M}$, e.g., $10^{5}$ Pa s, the MC viscosity can be as low as $10^{12}$ Pa s if $\phi = 50\% $. Considering the crystal-free, pure melt viscosity in the order of $10^2$ Pa s, the suspension (i.e., crystal-bearing melts) must contain solid crystals by 60-70\% to attain its viscosity in the order of $10^5$ Pa s (discussed in the earlier section). It means the pure melt percentage in the complex must be in the range of 15 to 20\%. The graphs (Figure \ref{Fig7}) show an inverse relation of the mush viscosity with melt suspension content; $\mu_{eff}$ becomes $10^{14}$ Pa s as $\phi$ decreases to 36\%, which corresponds to a pure melt fraction of 11-15\%. This melt fraction estimate would be further low if the polydispersity and polymodality factors were considered in the calculation. Similarly, an increase in $\mu_{M}$ (e.g., $10^{5}$ to $10^{7}$ Pa s) can yield the MC viscosity ($\mu_{eff}$) in the order of $10^{14}$ Pa s for $\phi = 40\%$ (Figure \ref{Fig7}). On the other hand, both $\mu_{M}$ and $\phi$ in the mush can increase to yield the same mush viscosity, i.e., $10^{14}$ Pa s for $\phi = 57\%$, and  $\mu_{M} = 10^{10}$ Pa s (Figure \ref{Fig7}). The two schematics in Figures \ref{Fig7}e and \ref{Fig7}f show sub-crustal melt bodies and magma conduits (i.e., MC region) with contrasting suspension characteristics at the two types of ridges. The mush complexes in faster ridges generally have melts with larger phenocrysts and groundmasses in larger volume fractions than those in slow-spreading ridges. Consequently, a higher effective viscosity of MC due to higher crystal content, polydispersity, and polymodality (Figures \ref{Fig7}a and \ref{Fig7}e), produces axial high topography in fast ridges. On the other hand, the opposite suspension characteristics sets in a low viscosity condition (Figures \ref{Fig7}a and \ref{Fig7}f) beneath slow ridges, which gives rise to axial flat topography. Melt contents in the MC can, however, fluctuate due to a number of factors, such as sub-crustal solidifications and numerous volcanic events in the process of new crust formation. A concerted operation of the following three processes: 1) mid-oceanic ridge eruptions, 2) sub-crustal solidifications \cite{Mandal2018}, and 3) continuous convective upwelling of partial melts \cite{Sarkar2014} can modulate the $\mu_{M}$ -- $\phi$ regression to maintain a steady-state µeff  condition (Figure \ref{Fig7}d) required for the stable axial topography.

\subsection{Axial topographic growth: mechanisms and their validation}

Several MOR models have attempted to integrate sub-ridge thermomechanical processes in the mantle-lithosphere, giving a spectrum of competing mechanisms, such as magmatic upwelling versus hydrothermal cooling \cite{Chen2022}, tectonic extension versus diking \cite{Buck2005}, fault-driven collapse versus isostatic compensation \cite{Escartin2008, Lin1990}, overpressure building versus  release of magma chambers \cite{Gudmundsson2012, Reverso2014}, and fluid convection versus matrix compation \cite{Katz2010}. The MOR model of our present concern invokes a sub-ridge mechanism of porous convection with synkinematic melting-solidification processes to describe the FSI mechanics. The convection-driven upwelling occurs at a depth of cessation of the adiabatic decompression melting. We use this threshold depth to introduce random temperature points (range $800^{\circ}$C to $1400^{\circ}$C) on a narrow region beneath the MOR axis (Figure S.3a). This random thermal perturbation (RTP) initiates the convection in the porous upper mantle, where the porous convective flow accompanies melting and solidifications in the sub-ridge shallow upper mantle and sub-crustal regions mediated an enthalpy transfer process. The flows are always geometrically asymmetric due to concerted effects of the RTP, porous convection \cite{Katz2010} and the intermixing of multiple convection cells and melting-solidification processes.\par

The convective flows induced in the MC develop normal stresses at its interface with the overlying crust, as modelled through a fluid-structure interaction (FSI). The effective viscosity of MC and its kinematic condition determines the magnitude of normal stresses transmitted to the overlying solid crust. The kinematic conditions of the MC are nominally transmitted to the overlying crust together with the dynamic conditions as an implementation of the Robin transmission condition in the FSI mechanism \cite{Badia2008}. In the FSI formulation, the shear stresses are generally excluded, considering that the differential velocity across the interface is small. However, for lower MC viscosities this factor can be significant due to the existence of strong relative motion. The stress transmission eventually results in deformations in the elastic crust to produce an axial topography. The effective viscosity of MC plays a critical role in controlling the magnitude of transmitted stresses that ultimately determine the vertical reliefs in the overlying elastic crust. Decreasing MC viscosities consequently results in a transition of high to flat axial topography. However, density takes the lead role to maintain flat topography at lower MC viscosities ($\leq10^{12}$ Pa s) (Figure \ref{Figure S5}).\par

\begin{figure}[ptbh]
\includegraphics[width=\textwidth]{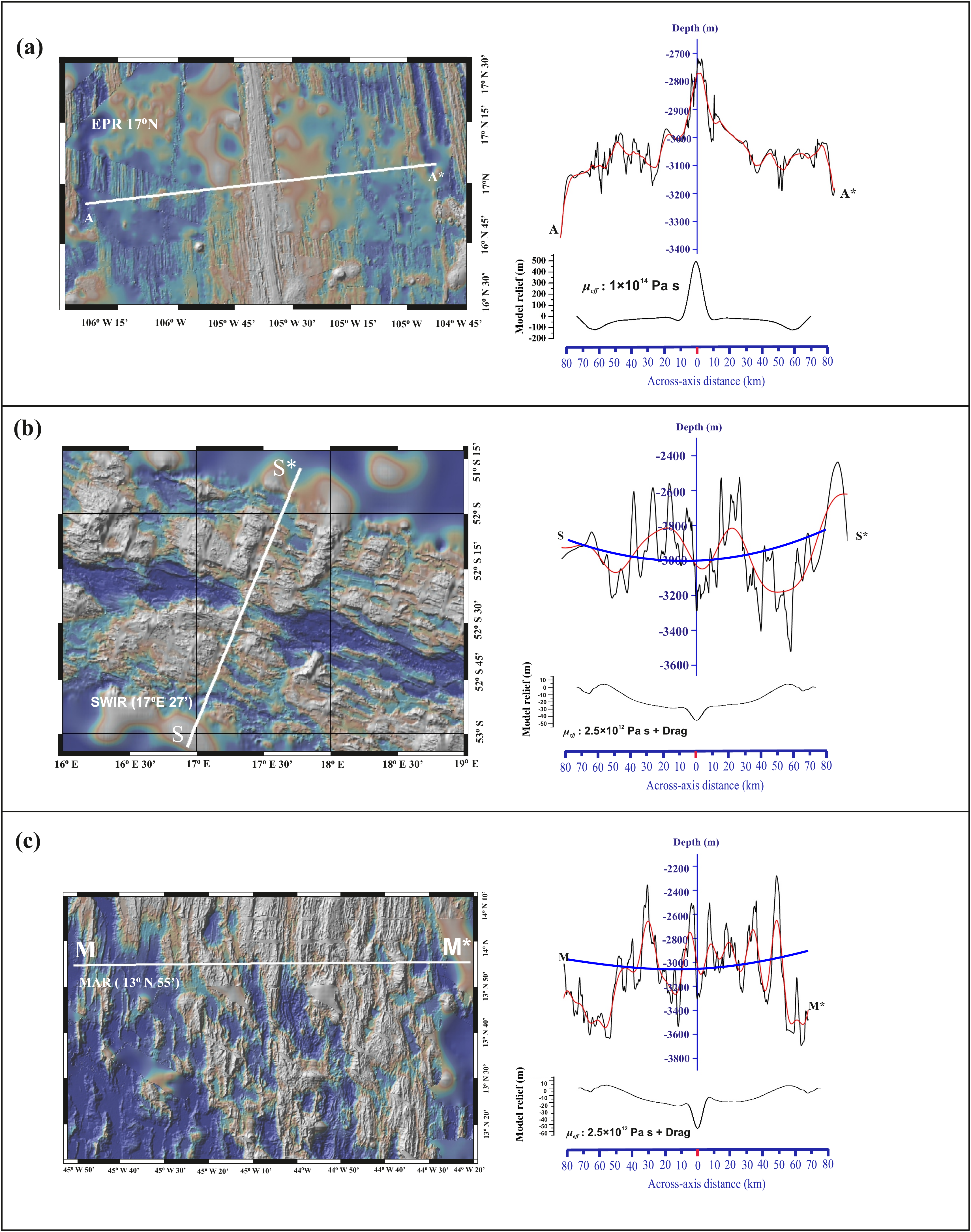}
\caption{Comparisons of the cross-axis topographic profiles between nature and model: (a) a high viscosity ($\mu_{eff} = 10^{14}$ Pa s) model versus EPR 17$^{\circ}$N. (b) a low-viscosity ($\mu_{eff} = 2.5\times10^{12}$ Pa s) model with drag force and SWIR 17$^{\circ}$27'E. (c) a low-viscosity ($\mu_{eff} = 2.5\times10^{12}$ Pa s) model with drag force and MAR 13$^{\circ}$55'N. Red lines show polynomial fits (higher-order) to the original natural data (in black). Blue lines indicate a second-degree polynomial fit for the SWIR and MAR. The EPR section displays a good match with the model topography [maximum ridge elevation (H): 400 m (model) and 500 m (nature), axial width (W): 40 km
(model) and 30 km (nature), similar ‘neck’s on both sides]. The SWIR section also shows a similarity with the model in their first-order axial topography, barring quantitative differences [H: - 40 m (model) and - 400 m (nature), W: 20 km (model) and 40 km (nature)]. The MAR
section matches fairly with the first-order model topography, but with differences in their magnitude [H: - 60 m (model) and - 600 m (nature), W: 15 km (model) and 10 km (nature)]. MOR data source: GeoMapApp (http://www.geomapapp.org/)/CC BY.}
\label{Fig8}
\end{figure}

We will now use some generic parameters of the natural MOR systems to test the validity of the FSI mechanism proposed in this study. The flow data of MC support the dynamic magma budget of mid-oceanic ridge calculated and validated with natural ridge processes in previous studies \cite{Mandal2018}. The melt budget suggests eruptible melts amount to $8-10\%$ of the total upwelling melt beneath MORs, which means, 3.7x$10^{6}\hspace{1mm} m^{3}/yr$  in a 500 km long ridge.  Secondly, the model presented here treats MC as a fluid region consisting of liquid mixtures of crystal-bearing melts and host rocks. Its viscosity analysis yields a value in the range $10^{12}$ to $10^{14}$ Pa s, which is comparable to that of lower-crustal magma bodies in MORs. For example, Chenevez \emph{et al.},\cite{Chenevez1998} estimated the viscosity of gabbroic mushes within the axial magma chamber as $10^{15}$ Pa s from Oman Ophiolites. In addition, McKenzie \cite{McKENZIE1984} considered the effective viscosity of melt-bearing matrix in the order of $\sim 10^{15}$ Pa s. On the other hand, experiments have shown viscosity in the order of $10^{11}$ Pa s for lower crusts containing melts by 20-25\% \cite{Picard2013}. Similarly, Fontaine \emph{et al.}\cite{Fontaine2017} estimated an effective viscosity of $10^{13}$ Pa s for sub-crustal regions in the melt-rich fast spreading ridges showing axial highs, as produced in our model (Figure \ref{Fig6}a). The spectrum of melt percentages at MC, as predicted from the present rheological calculations (30-80\%, see Figure \ref{Fig2}a-d), is supported by earlier estimates. The strain rates in the MC ($10^{-3} s^{-1}$ to $10^{-11} s^{-1}$) with a median value of $10^{-5} s^{-1}$ also agree well with those recorded in natural MOR systems. Our FSI model shows that the timescale of mid-oceanic ridge processes to stabilize (Figure \ref{Fig5}) is 6 Myr, which is comparable to those reported in the literature \cite{Gerya2013}.

\subsection{Axial topography: a model versus nature comparison}

\begin{figure}[ptbh]
\includegraphics[width=\textwidth]{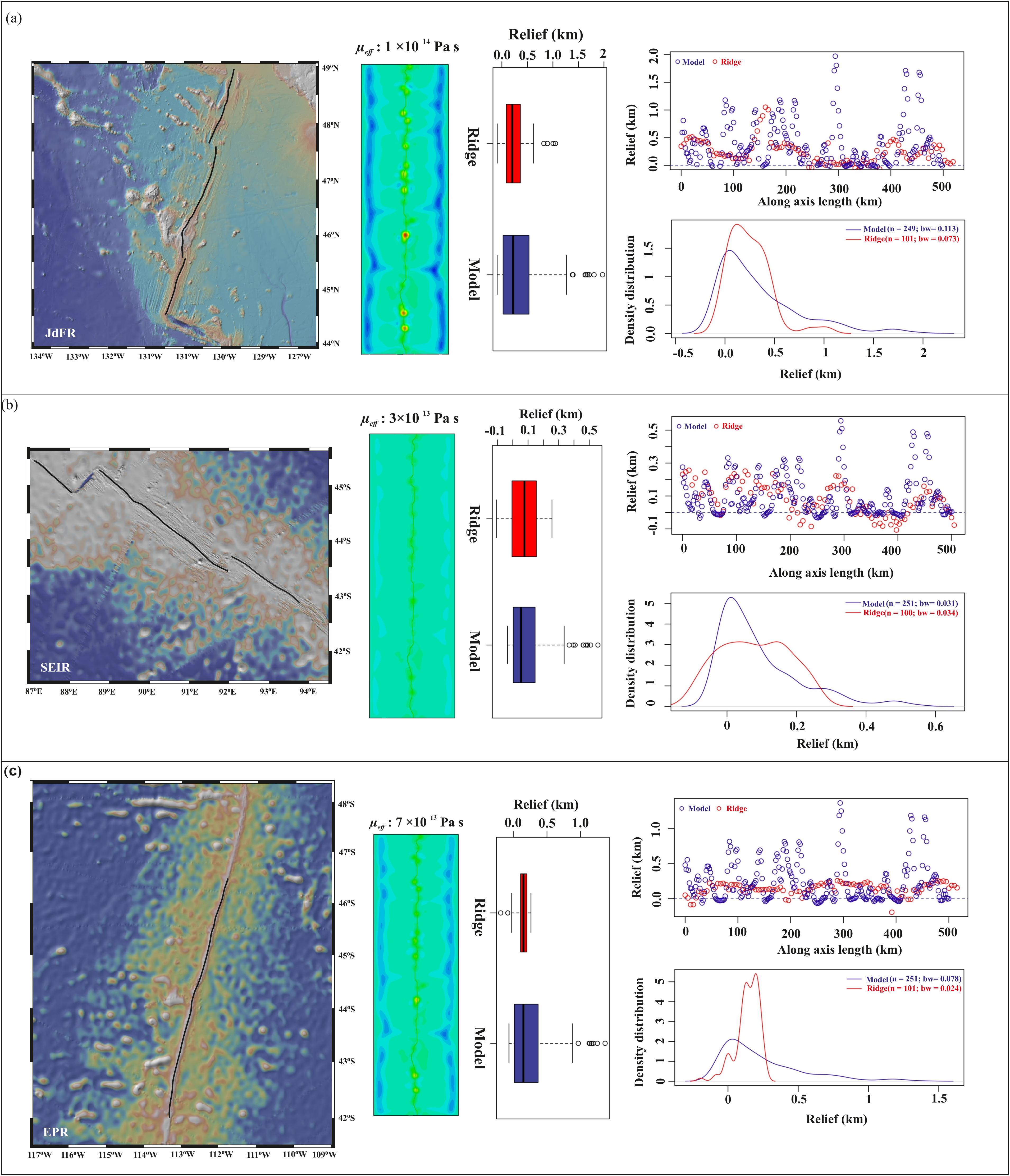}
\caption{Comparison between natural and model along-axis topography: (a) A high viscosity ($\mu_{eff} = 10^{14}$ Pa s) model and JdFR (44$^{\circ}$30’N to 49$^{\circ}$N). (b) A moderate viscosity
($\mu_{eff} = 3\times10^{13}$ Pa s) model and eastern SEIR (87$^{\circ}$ 30’E to 93$^{\circ}$30’E).(c) A moderately high viscosity ($\mu_{eff} = 7\times10^{13}$ Pa s) model and EPR (42$^{\circ}$S – 46$^{\circ}$30’S). The box plots, scatter plots, and density distributions are also shown, along with the natural ridge bathymetry and the evolved model ridge axis elevations. For JdFR, the along-axis profiles show matching topography with the model [median relief : 0.2043 km (natural) and 0.22 km (model); density peak : 0.12 km (natural) and 0.05 km (model)]; SEIR and EPR profiles are also in good agreement with the model topography: for SEIR, median relief : 0.076 km (natural) and 0.054 km (model); density peak : 0.05 km (natural) and 0.02 km (model); and for EPR, median relief : 0.15 km (natural) and 0.15 km (model); density peak : 0.22 km (natural) and
0.04 km (model). Natural data source: GeoMapApp (http://www.geomapapp.org/)/CC BY.}
\label{Fig9}
\end{figure}

In the quantitative analysis of axial topography the median relief of natural ridge systems has been considered to estimate the vertical anomaly with respect to the average seafloor depth (2600 m, \cite{Searle2013}). We chose an across-axis section of the EPR at $17^{\circ}$N \cite{Lin1989} to compare its long wave axial-high topography with those obtained from our model. The $17^{\circ}$N section displays a first-order characteristic topography consisting of a sharp axial high (maximum elevation: H $\sim$ 400~m, axial width: W $\sim$ 40~km), flanked by a symmetric pair of flat regions (width $\sim$ 60~km), and narrow, weak depression zones away from the high. The axial topography shows a good match with that produced in the model simulation ($\mu_{eff} = 10^{14}$ Pa s, H $\sim$ 500~m and W $\sim$ 30~km) (Figures \ref{Fig8}a). We also validated along-axis model topographic patterns with the available natural data. Figure~\ref{Fig9}a shows a comparative analysis of the Juan de Fuca (JdF) ridge-segment ($44^{\circ}30'$N to $49^{\circ}$ N) topography and the $\mu_{eff} = 10^{14}$ Pa s model axis relief (at 7 Myr model run-time). The JdF ridge includes a seamount, and six major segments form reliefs with their median close to the model value (220 m). The model and the natural ridge systems show remarkable similarity in terms of the relief density distribution and scatter (Figure~\ref{Fig9}a). Geologically, the JdF ridge (JdFR) with moderate spreading rates (60 mm/yr) receives lateral magma supply from the neighbouring Cobb hotspot and axial seamounts. Enhanced fractional crystallization with efficient cooling \cite{Chadwick2005} away from the hotspot regions thus produces a viscous magma-enriched sub-crustal melt-rich system, which in turn facilitates the growth of axial high topography, as predicted from high-viscosity FSI model (Figure \ref{Fig9}a). Additional notable features of JdFR are: 1) the axial seamount does not significantly differ from the adjoining parts of the axial ridge segment in terms of the crystal content of their magmas \cite{Chadwick2005}, and 2) the excess melt volume is compensated by forming a thick crust. However, there is a possibility of narrow, focused upward magma flux to the axial seamount’s base, resulting in enhancement of the normal flux in JdFR by three times \cite{West2003}, as reflected from higher upwelling velocity/strain rates in this region. The ridge seamount thus represents a local feature to enhance the vertical strain rates over the common viscous behaviour of the underlying mush and give rise to topographic characteristics observed in the corresponding model, where the positive relief is larger than the maxima by 60\%, and ten times the median value (Figure \ref{Fig9}a).\par

We compared across-axis model topographic profiles with two sets of nearly flat topography from ultraslow SWIR and slow-spreading MAR extrapolated directly from the GeoMap database. These two ridges exhibit predominantly rifted valley topography. We thus chose two narrow segments, where rifting is not a dominant ridge process, as indicated by thick crust, but they show weak axial valleys or flat axial topography. 
A topographic profile from the low-viscosity ($\mu_{eff} = 2.5\times10^{12}$ Pa s) MC model compares well with the first-order valley geometry ($17^{\circ}27'$E, \cite{Grindlay1998}) in the SWIR, when the off-axis depressions due to extensional faulting are excluded (Figure \ref{Fig8}b). In the case of MAR, another profile of the same model run grossly reproduces the ridge profile ($13^{\circ}55'$ N, \cite{Mallows2012}), which consists of a narrow, shallow valley at the ridge flanked by flat off-axis stretches (Figure \ref{Fig8}c). However, there are large differences in the magnitudes of axial depression topography between the natural settings and the corresponding models (e.g., $\sim$30 m in model vs. $\sim$100 m in SWIR, $\sim$50 m in model vs. $\sim150$ m in MAR, $13^{\circ}55'$ N) but they show a first-order similarity in their axial zone topography, e.g., across-axis width ($\sim$80 km) of the gentle axial depressions. However, the higher-order off-axis topographic elements in model and nature do not perfectly match with one another (Figure \ref{Fig8}c). These higher-order mismatches perhaps result from strong effects of tectonic (tensile) stresses, as compared to relatively weak rheological effects of the underlying mush complex.\par

We extended our model validation with a number of natural along-axis topographic profiles from SEIR ($87^{\circ}30'$ E to $93^{\circ}30'$ E). These profiles show a marked similarity in their relief patterns (Figure \ref{Fig9}b) with those in the model run for moderate effective MC viscosity ($\mu_{eff} = 3\times10^{13}$ Pa s at 7 Myr run time). At the western portion the SEIR has a significant influence of melt plumes \cite{Baker2014}, and developed a first-order transform discontinuity and a prominent overlapping spreading centre. Overall, the ridge, deepening towards the east \cite{Sempere1997}, displays along-axis roughness of its relief fairly in agreement with our model. The western part of SEIR, extending up to $90^{\circ}$E receives mantle-derived melts from the Kerguelen-Heard hotspot, as suggested by the ${}^{87}Sr/~{}^{86}Sr$ ratio enrichment. Alternatively, there is a possibility for greater availability of partial melts due to a greater mean depth of melting, which is correlated with the He isotope ratio peak at $88^{\circ}$E (see Mahoney \emph{et al.}\cite{MAHONEY2002} and references therein). Both the cases can give rise to a condition of high magma percentage and low magma viscosity in the portion of SEIR of our present concern, which might retain $\mu_{eff}$ at moderate values ($\sim 3 \times 10^{13}$ Pa s), as derived from the viscosity calculations (Figure \ref{Fig7}b).\par

We support our model interpretations with a positive correlation of the along-axis model topography with the southern part of the EPR and the northern segment of the PAR ($42^{\circ}$S – $46^{\circ}30'$S) (Figure \ref{Fig9}b). The latter is thought to be stable for more than 50 Myr \cite{Rowley2016}. Its median relief matches well with that of a high ($\mu_{eff} = 7\times10^{13}$ Pa s) model. On the other hand, the EPR ridge segment shows a lower relief variance than the model (Figure \ref{Fig9}c).The overall topographic parity allows us to predict the viscosity of melt-rich sub-crustal region in the order of $10^{13}$ Pa s for this particular ridge segment. Applying our two-phase viscosity model, we suggest that although this region is rich in melt content, it gains relatively high viscosity due to a large volume fraction of solid crystals in the melt suspensions. The relatively lower along-axis variance in the EPR ridge topography, as compared to that in the corresponding model, results from a number of possible factors, such as longitudinal stability of the ridge position, continuous deep-seated upwelling, and overwhelming viscous magmatic control \cite{Rowley2016}. A rate balance between the melt supply and crystallization can account for a steady-state effective viscosity of the underlying melt-bearing regions to sustain such stable ridge-axis topography (see Figure \ref{Fig7}d).\par

\begin{figure}[ptbh]
\includegraphics[width=\textwidth]{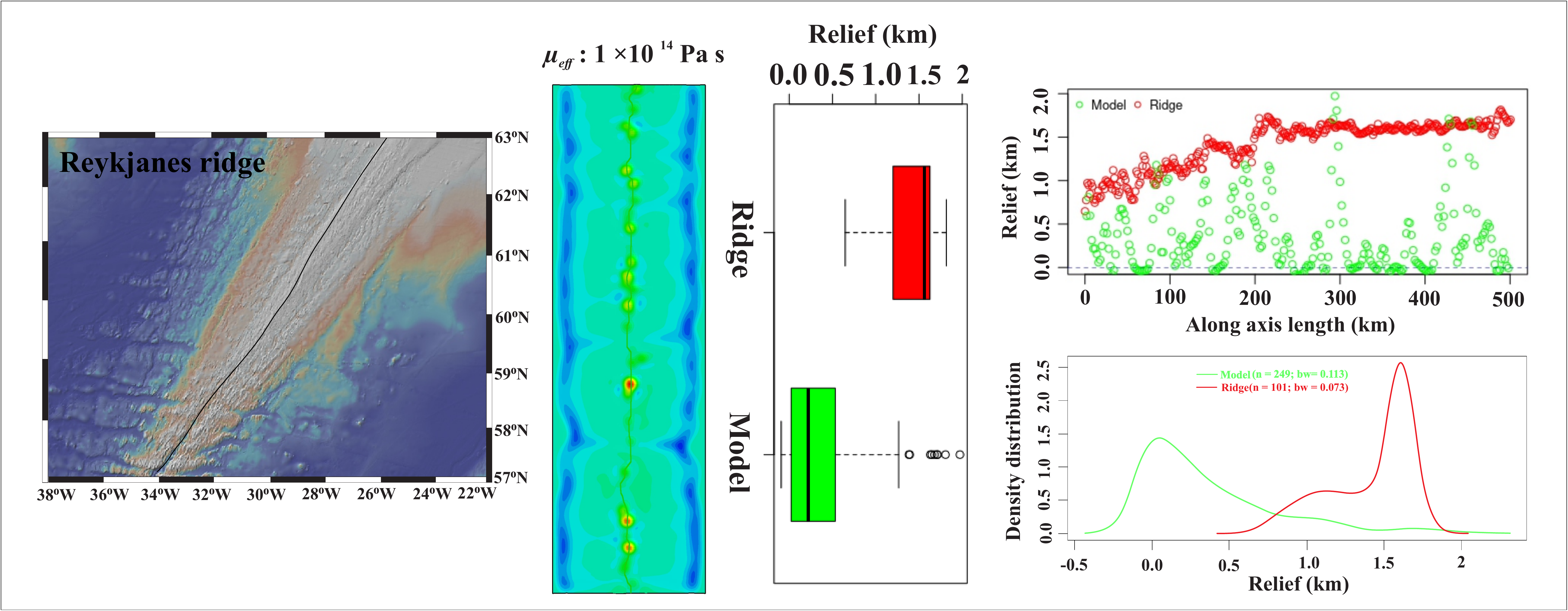}
\caption{Comparison between natural and model along-axis topography of Reykajanes Ridge (44$^{\circ}$30’N to 49$^{\circ}$ N). A high viscosity model ($\mu_{eff}  = 10^{14}$ Pa s) is shown in the same panel producing Axial high topography. The box plots, scatter plots, and density distributions are also shown, along with the natural ridge bathymetry and the evolved model ridge axis elevations. Natural data source: GeoMapApp (http://www.geomapapp.org/)/CC BY}
\label{Figure S6}
\end{figure}

This discussion leads us to suggest the following. The magma-rich EPR has retained a high-viscosity condition of the melt-bearing sub-crustal regions to produce narrow axial high, as produced in our simulation with $\mu_{eff} = 5\times10^{13} - 10^{14}$ Pa s. Hotspot fed and rapidly cooling melts in the JDFR has a sub-ridge MC with the highest viscosity ($\mu_{eff} = 10^{14}$ Pa s), and their produced an axial high elevation comparable to that in the model. On the other hand, the western SEIR is rich in moderately viscous melt suspensions but becomes melt-poor, resulting in deep rifted axial valley topography in the eastward direction \cite{Sempere1997, Baker2014}. The analysis suggests a moderate sub-crustal viscosity ($\mu_{eff}$) has formed axial high topography in western SEIR, which agrees well with the simulation result for $\mu_{eff} = 1-5\times10^{13}$ Pa s. Our FSI model for $\mu_{eff} \sim 10^{12}$ Pa s points to an appreciable mismatch on the along-axis model topography with those observed in the magma poor MAR and SWIR (both slow-spreading), albeit showing a reasonable match with the across-axis first-order curvatures in their non-rifted segments. We suggest that the MAR and SWIR topography are not entirely controlled by the rheological setting of their sub-crustal and lower crustal mush complexes. This mismatch indicates the possibility of tensile stress regimes to govern the axis topography where the flow-driven stresses at the base in case of slow-spreading ridges become relatively weak due to low-viscosity condition in the underlying mantle (e.g., Lin and Parmentier\cite{Lin1989}). Our model also shows a weak match of the along-axis model topography with the Reykjanes ridge topography where the relief is significantly higher than the model relief even in high-viscosity ($\mu_{eff} \sim 10^{14}$ Pa s) simulations (Figure \ref{Figure S6}). It is noteworthy that the 600 km long slow (2 cm/year full spreading rate) Reykjanes ridge ($57.9^{\circ}$ N to $62.10^{\circ}$ N) is thought to have evolved under the influence of Iceland mantle plume, as evident from its large oblique spreading characteristics (280 from the spreading normal) and its V shaped plan view \cite{Searle1998, White1995}. This might be the reason for the topographic mismatch.\par

Axial relief in the present model correlates positively with crystal contents, polymodality and polydispersity in the MC that enhances the melt suspension viscosity, and in turn, the effective viscosity of the MC. Faster spreading ridges show crystallization at shallower depths \cite{Wanless2012} and they also undergo greater mixing of their crystal phases (olivine, plagioclase and clinopyroxene) of different sizes and shapes at subcrustal / lower crustal MC \cite{Lissenberg2019}. The predominance of crystal suspensions sets in a high-viscosity rheological condition that explains the axial highs at faster spreading segments. In contrast, crystallization at slow spreading ridges typically occurs at greater depths \cite{Herzberg2004}, allowing the melts to transport through melt channels, but loosing heavier (olivine) larger crystal in their pathways due to slow ascent velocities \cite{Lange2013}. In effect, slow spreading ridges are likely to produce MCs with low crystal contents and lower polymodality and polydispersity that result in setting up a low-viscosity setting and weak normal stress transfer in the topographic process.

\subsection{Model limitations}

The present FSI model is designed to study the sole role of the viscosity of melt-rich regions in controlling the axial topography of mid-ocean ridges. However, as discussed in the Introduction, several other factors, e.g., the diking parameter \cite{Liu2018}, can influence the ridge topography. A large number of studies have recognized the spreading rate as a potential factor to modulate the axial high versus valley development. The exclusion of this factor obviously imposes a limitation on our modelling. However, this study sheds light in a new direction, showing that the viscosity changes of sub-crustal melt regions by $10^{1}$ – $10^{2}$ order can alone bring a transition from flat to axial high topography in a MOR under the same spreading rate. Secondly, MORs generally undergo extensional faulting in the uppermost brittle crustal layer \cite{Buck2005}, which contributes to the development of high-order ocean floor morphology, such ridge parallel hills at MORs. Also, extensional stresses play a dominant role in forming axial valleys \cite{Lin1989}. Our model excludes such tectonic stress regimes at MOR and brittle failure in the elastic solid top layer as we focus on longer wavelength topography. The gradient in across-axis lithospheric thickness variation might have an additional influence in the axial topographic development. However, this factor excluded in this study to find independently the effects of sub-crustal mush complexes (MC) on the two end-members: flat and high axial topography.

\section{Conclusions}

1) One-way Fluid-Structure coupling between a sub-crustal melt accumulation zone and the overlying solid elastic crust, has been implemented with the framework of a computational fluid dynamics modelling of convective heat and mass transfer. The model results demonstrate that the effective viscosity of the melt-rich zones can play a critical role in modulating the axial high versus flat topography.
2)  We claim that the mush complex (MC) dynamics involving crystallization in its melt suspensions at the lithospheric base can largely govern the ridge axis topography.
3)  A complete description of the MOR mechanical setting demands viscosity analysis on two scales- one at magma body/conduit scale, which is tackled by utilizing suspension theory, and mush scale, which is dealt with a modified Arrhenius equation.
5) The effective viscosity of MC varying in the range $10^{12}$ Pa s to $10^{14}$ Pa s produces a full spectrum of the non-rifted axial high to flat (1.27 km to - 0.06 km) topography. Typical axial highs form in the viscosity range of $10^{13}$ Pa s - $10^{14}$ Pa s, whereas axial lows in the viscosity range of $10^{12}$ Pa s - $5\times10^{12}$ Pa s. 
6) The onset of relative vertical displacements in central axial regions occurs at the time of upwelling melt-bearing mushy materials to interact with the overlying crust. The process forms a stable topography on a time scale of $\sim$ 6 to 7 Myr, characterized by central axial highs and off-axis depressions on their flanks.
7) This viscosity based new model explains the following characteristics of natural MORs: a) axial high topography in melt-rich ridge systems (e.g., EPR) and first-order axial valley in melt-poor ridges (e.g., SWIR and MAR), b) transformation of ridge topography due to drastic changes in subcrustal magma constituency (e.g., SEIR), c) axial seamount as a location of high upwelling rates (e.g., JdFR) and d) stability of axial topography in large temporal and spatial settings (e.g., Southern EPR) as an outcome of the competing factors, such as viscosity and volume percentage of crystal-bearing melt suspensions that maintain the effective viscosity of mush complex almost at a constant level.
8) Our FSI modelling constrains the viscosities of sub-crustal mushy regions in the following MOR systems: $10^{14}$ Pa s for JdFR, $5\times10^{13}$ - $10^{14}$ Pa s for EPR, 1 - $5\times10^{13}$ Pa s for western SEIR.

\bibliographystyle{acm}
\bibliography{main}
\end{document}